\documentclass[amsmath, amssymb, superscriptaddress]{revtex4-2}
\pdfoutput=1
\usepackage{graphicx}
\usepackage{siunitx}

\usepackage[
pdftitle={Bloch points in nanostrips},
pdfauthor={Martin Lang, Marijan Beg, Ondrej Hovorka, and Hans Fangohr},
colorlinks=True,linkcolor=black,citecolor=black
]{hyperref}

\newcommand{\nm}{\ensuremath{\,\mathrm{nm}}}
\newcommand{\BP}[1][ ]{Bloch{#1}point}

\begin{document}

\title{Bloch points in nanostrips}

\author{Martin Lang}
\email{m.lang@soton.ac.uk}
\affiliation{Faculty of Engineering and Physical Sciences, University of
  Southampton, Southampton, SO17 1BJ, United Kingdom} 
\affiliation{Max Planck Institute for the Structure and Dynamics of Matter,
  Luruper Chaussee 149, 22761 Hamburg, Germany}

\author{Marijan Beg}
\affiliation{Faculty of Engineering and Physical Sciences, University of
  Southampton, Southampton, SO17 1BJ, United Kingdom}
\affiliation{Department of Earth Science and Engineering, Imperial College
  London, London SW7 2AZ, United Kingdom}

\author{Ondrej Hovorka}
\affiliation{Faculty of Engineering and Physical Sciences, University of
  Southampton, Southampton, SO17 1BJ, United Kingdom} 

\author{Hans Fangohr}
\affiliation{Faculty of Engineering and Physical Sciences, University of
  Southampton, Southampton, SO17 1BJ, United Kingdom}
\affiliation{Max Planck Institute for the Structure and Dynamics of Matter,
  Luruper Chaussee 149, 22761 Hamburg, Germany}
\affiliation{Center for Free-Electron Laser Science, Luruper Chaussee 149, 22761
  Hamburg, Germany}

\begin{abstract}
  Complex magnetic materials hosting topologically non-trivial particle-like
objects such as skyrmions are under intensive research and could fundamentally
change the way we store and process data. One important class of materials are
helimagnetic materials with Dzyaloshinskii-Moriya interaction. Recently, it was
demonstrated that nanodisks consisting of two layers with opposite chirality can
host a single stable Bloch point of two different types at the interface between
the layers. Using micromagnetic simulations we show that FeGe nanostrips
consisting of two layers with opposite chirality can host multiple coexisting
Bloch points in an arbitrary combination of the two different types. We show
that the number of Bloch points that can simultaneously coexist depends on the
strip geometry and the type of the individual Bloch points. Our simulation
results allow us to predict strip geometries suitable for an arbitrary number of
Bloch points. We show an example of an 80-Bloch-point configuration verifying the
prediction.
\end{abstract}

\maketitle

Magnetic quasiparticles with non-trivial topology~\cite{Wang2021}, such as
vortices and skyrmions, are of great fundamental interest and could play an
important role in novel technological applications~\cite{Gobel2021}. One of the
quasiparticles is the Bloch point: a single point of vanishing
magnetisation~\cite{Feldtkeller1965, Doring1968}. A stable and manipulable \BP{}
of two different types was predicted in helimagnetic two-layer
nanodisks~\cite{Beg2019} where the two layers have opposite chirality, and the
\BP{} nucleates at the interface between the layers.

In this work, we demonstrate that two-layer nanostrips can host multiple \BP{}s
in an arbitrary combination of different \BP[-]{} types. Using finite-difference
micromagnetic simulations, we explore the parameter space of strip geometry to
understand for which geometry constraints such magnetisation field
configurations can be realised. The \BP{} as a (meta)-stable topological
excitation and quasiparticle opens up new avenues of fundamental research. In
particular, we can now start to investigate individual and collective behaviour
of \BP{}s towards discovering \BP[-]{}-based meta-materials. We conclude our
study with a demonstration of encoding a 10-byte string using 80 \BP{}s: we
identify one \BP[-]{} type with the binary ``\texttt{1}'' and the other type
with the binary ``\texttt{0}'' to encode and store the equivalent of an 80-bit
long sequence.

\begin{figure}
  \includegraphics[width=\linewidth]{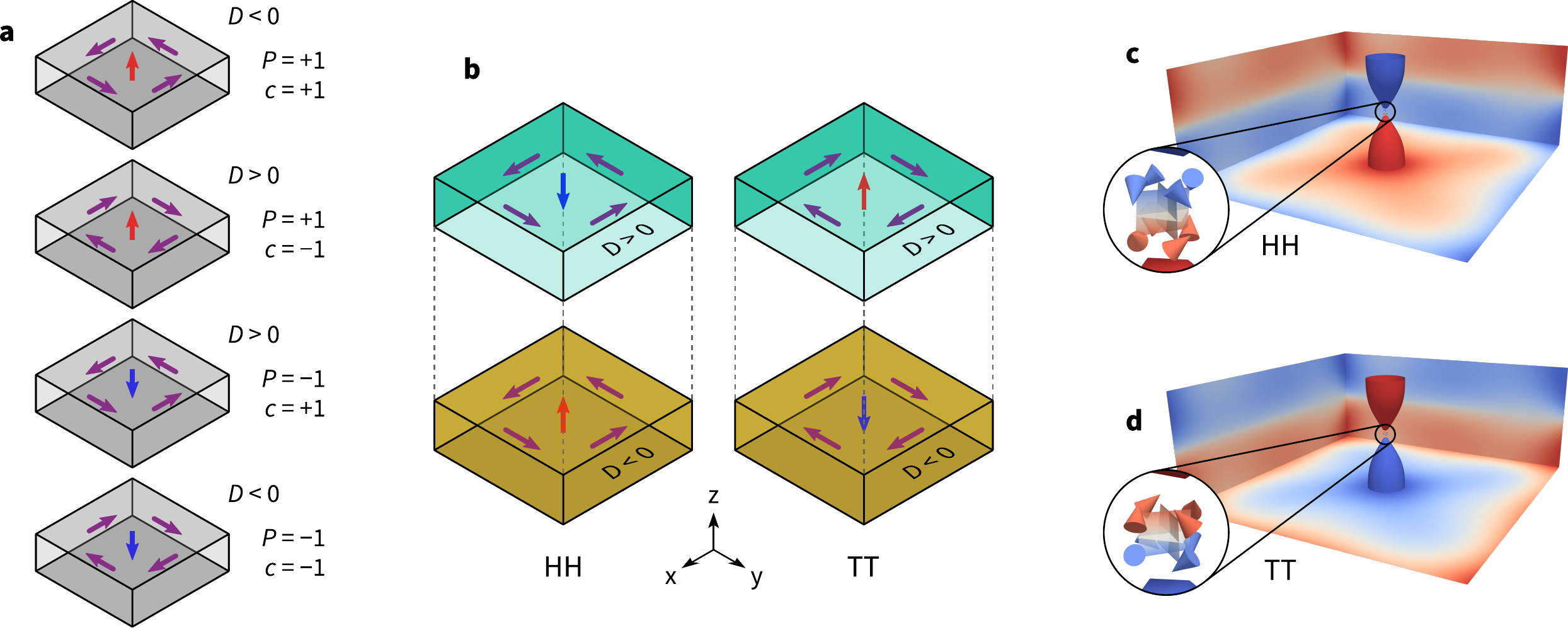}
  \caption{\label{fig:bp-concept}(a) In a single layer of magnetic material four
    different vortices, with polarisation $P=\pm 1$ and circularity $c=\pm 1$,
    can form as a consequence of the competition between exchange energy and
    demagnetisation. Adding DMI couples circularity and polarisation. (b) By
    stacking two layers with opposite sign of the DM energy constant $D$, a
    \BP{} can be stabilised. The \BP{} can be of type head-to-head (left) or
    tail-to-tail (right). In the figure, the two layers are, for better clarity,
    separated in $z$ direction as indicated by the grey dashed lines. (c, d)
    Simulation result for a single head-to-head (c) and tail-to-tail (d) Bloch
    point. The isosurfaces (of paraboloidal-like shape) near the centre show
    $m_{z}=\pm0.9$, colour indicates the $z$ component. They are convenient to
    locate the \BP{} that is situated between them. The insets show three
    isosurfaces for $m_{x} = 0$, $m_{y} = 0$, and $m_{z} = 0$, respectively. The
    \BP{} is located at the intersection of the three isosurfaces where the
    magnetisation vanishes. The cones in the insets indicate the magnetisation
    directions in the eight discretisation cells surrounding the Bloch point.}
\end{figure}

The concept of a \BP{} in a two-layer system~\cite{Beg2019} is explained in
Fig.~\ref{fig:bp-concept}, where we start from vortex configurations in
single-layer materials, and then stack them on top of each other to obtain the
\BP[-]{} configuration of the magnetisation field. Figure~\ref{fig:bp-concept}a
shows schematically the four possible vortex configurations we can encounter in
a thin layer of ferromagnetic material due to the competition between
ferromagnetic exchange and demagnetisation energy. The vortex core (polarisation
$P$), pointing in the out-of-plane ($z$) direction, can either be pointing to
$+z$ ($P=+1$) or $-z$ ($P=-1$) direction, and -- independently -- the
circularity $c$ of the magnetisation around the vortex core can either be
clock-wise~($c=-1$) or counter-clock-wise~($c=+1$)~\cite{Behncke2018}. By adding
the Dzyaloshinskii-Moriya (DM) interaction to the system, the relation between
polarisation and circularity is fixed through the chirality, \emph{i.e.} the
sign of the DM interaction strength $D$: for a given $D$ only two of the four
vortex realisations are energetically favourable.

Figure~\ref{fig:bp-concept}b shows how a \BP[-]{} configuration can be realised by
stacking vortex configurations with the same circularity and opposite
polarisation on top of each other. The \BP{} emerges at the interface between
the two vortex cores of opposite polarisation, and the top- and bottom-layer
materials must have opposite signs of $D$ to stabilise the \BP[-]{} configuration.
Despite the high exchange energy density at the \BP{}, the configuration is
stabilised by the exchange coupling across the comparatively large interface
area between the layers that ensures the same circularity in both layers. Two
different types of \BP{}s can be realised with the magnetisation vectors of the
vortex cores pointing either inwards (head-to-head \BP{},
Fig.~\ref{fig:bp-concept}b~HH) or outwards (tail-to-tail \BP{},
Fig.~\ref{fig:bp-concept}b~TT).

Figure~\ref{fig:bp-concept}c shows the magnetisation vector field for a single
head-to-head Bloch point from our finite-difference micromagnetic simulations,
and the colour represents the $z$ component of the magnetisation.
Figure~\ref{fig:bp-concept}d shows the corresponding plot for a tail-to-tail
Bloch point. In the simulation, the \BP{} forms at the centre of eight
discretisation cells as shown in the insets of Fig.~\ref{fig:bp-concept}c and~d
where the magnetisation of the discretisation cells is visualised by the cones.
The \BP{} is located at the intersection of the three isosurfaces $m_{x, y, z} =
0$ where the magnetisation vanishes.

The paper is organised as follows. First, we discuss two different
configurations containing two Bloch points each in Sec.~\ref{subsec:two-bps}. We
distinguish between configurations of two same-type \BP{}s and two opposite-type
\BP{}s. These two pairs of \BP{}s exhibit all features of a configuration
containing multiple \BP{}s. In Sec.~\ref{subsec:parameter-space}, we demonstrate
that multiple \BP{}s can coexist in rectangular two-layer nanostrips using
micromagnetic simulations. We find that all possible sequences of head-to-head
and tail-to-tail \BP{}s can be realised. Different combinations have different
energy densities depending on the number of neighbouring \BP{}s of the same
type. We focus on systems containing up to eight \BP{}s initially. Based on our
results, we can predict a suitable strip geometry for an arbitrary number of
\BP{}s. To verify this prediction we present one example configuration
containing 80 \BP{}s in Sec.~\ref{subsec:80-bp}.

\section{Results}
\label{sec:results}
\subsection{Two Bloch points}
\label{subsec:two-bps}
\begin{figure}
  \includegraphics[width=.95\linewidth]{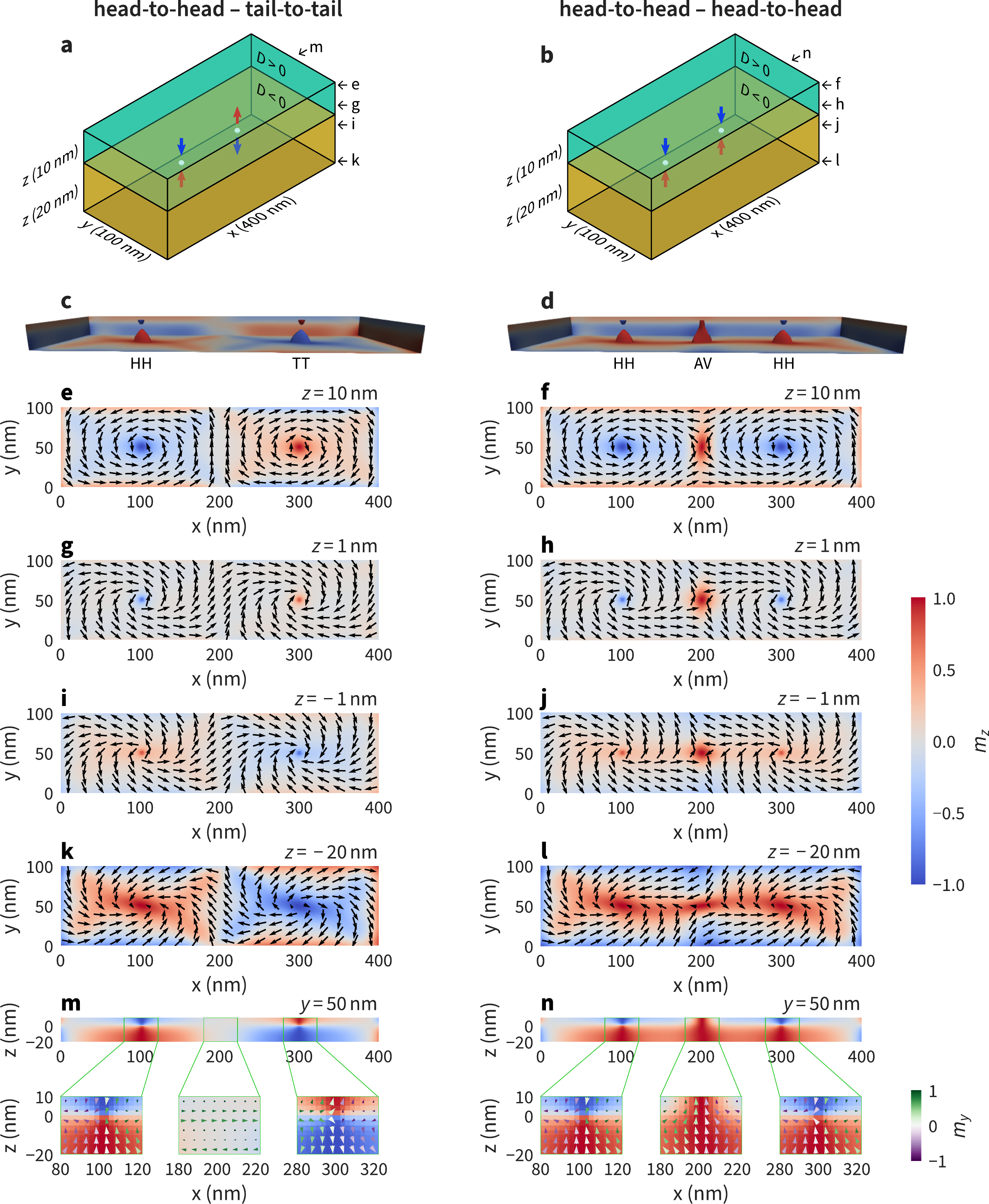}
  \caption{\label{fig:two-bp-magnetisation}Magnetisation profile of the two
    fundamentally different configurations containing two \BP{}s: opposite-type
    \BP{}s (head-to-head and tail-to-tail, left column) and same-type \BP{}s
    (head-to-head and head-to-head, right column). The 3D renderings in (c) and
    (d) show isosurfaces for $m_{z} = \pm 0.9$, colour indicates the
    $z$~component. Several different cut planes in $xy$ and $xz$ are shown to
    reveal the three-dimensional structure of the \BP{}s forming at the
    interface ($z=0\,\mathrm{nm}$). For the $xz$ plane (subfigures m and n) we
    also show enlarged plots around the \BP{} and antivortex position. The cones
    in (m) and (n) are coloured according to their $m_y$ component, as indicated
    by the small colour bar in the right bottom corner of the figure.}
\end{figure}

A pair of neighbouring \BP{}s in multi-\BP[-]{} configurations can occur in two
fundamentally different combinations. The \BP{}s can either be of the same type,
for example: a head-to-head (HH) \BP{} next to another HH \BP{} (HH-HH) as
shown in the right column of Fig.~\ref{fig:two-bp-magnetisation}. Alternatively,
they can be of opposite type, for example a HH \BP{} next to a tail-to-tail (TT)
\BP{} (HH-TT) as shown in the left column of
Fig.~\ref{fig:two-bp-magnetisation}.

The topmost row (Fig.~\ref{fig:two-bp-magnetisation}a, b) shows a schematic
drawing of the nanostrip highlighting the two-layer structure and the geometry.
Additionally, position and type of the \BP{}s visible in the simulations are
indicated with arrows, where the colour of the arrows encodes the $z$ component
of the magnetisation (red: $+z$, blue: $-z$).

Figure~\ref{fig:two-bp-magnetisation}c and d show 3D renderings of the
simulation results. The isosurfaces show $m_{z}=\pm 0.9$ ($\mathbf{m}$ is the
normalised magnetisation), colour indicates the $z$ component. The isosurfaces
above/below the \BP{} have a paraboloidal-like shape pointing towards the Bloch
point (similar to the single-Bloch-point simulation results in
Fig.~\ref{fig:bp-concept}c, d). The \BP{} itself is not directly visible in this
visualisation. The configuration in Fig.~\ref{fig:two-bp-magnetisation}d
additionally contains one antivortex between the two Bloch points. The $m_{z} =
0.9$ isosurface of the antivortex extends throughout the whole thickness ($z$
direction) of the two-layer system. The antivortex core shrinks towards the top
sample boundary.

To reveal the full three-dimensional structure of the magnetisation field
surrounding the \BP{}s the magnetisation of each configuration is plotted in
five different cut planes for each column (as indicated in the schematic
drawings Fig.~\ref{fig:two-bp-magnetisation}a,~b). Four different cut planes
show the magnetisation in the $xy$ plane
(Fig.~\ref{fig:two-bp-magnetisation}e~--~l), at the top sample boundary
($z=10\,\mathrm{nm}$) in Fig.~\ref{fig:two-bp-magnetisation}e and~f, above the
interface ($z=1\,\mathrm{nm}$) in Fig.~\ref{fig:two-bp-magnetisation}g and~h,
below the interface ($z=-1\,\mathrm{nm}$) in
Fig.~\ref{fig:two-bp-magnetisation}i and~j, and at the bottom sample boundary
($z=-20\,\mathrm{nm}$) in Fig.~\ref{fig:two-bp-magnetisation}k and~l. Colour
encodes the $z$ component of the magnetisation vector field, arrows the
in-plane component.

Figure~\ref{fig:two-bp-magnetisation}m and~n show the $z$ component of the
magnetisation in an $xz$ cut plane going through the \BP{}s at
$y=50\,\mathrm{nm}$. Magnified subplots show the full magnetisation around the
\BP[-]{} positions and in the centre region between the two \BP{}s. The colour
of the cones in the magnified areas encodes the $y$ component of the
magnetisation vector field.

The results shown in Fig.~\ref{fig:two-bp-magnetisation} show that \BP{}s form
at $x\approx 100\,\mathrm{nm}$ and $x\approx 300\,\mathrm{nm}$ in both cases,
\emph{i.e.} in both columns. \BP[-]{} pairs of the opposite type
(Fig.~\ref{fig:two-bp-magnetisation}, left column) show opposite circularity of
the magnetisation within the $xy$ plane around the \BP[-]{} cores. In this case,
the in-plane magnetisation ($xy$ component) between the two \BP{}s (from
$x\approx 100\,\mathrm{nm}$ to $x\approx 300\,\mathrm{nm}$) shows a smooth
transition from one \BP{} to the other. In contrast, an additional antivortex
forms between two same-type \BP{}s (Fig.~\ref{fig:two-bp-magnetisation}, right
column) at $x\approx 200\,\mathrm{nm}$ to mediate between the incompatible
magnetisation configurations that originate from the \BP{}s. Differing from the
magnetisation of the \BP[-]{} cores, the magnetisation of the antivortex core
(at $x \approx 200\,\mathrm{nm}$) does not change significantly along the $z$
direction (inset in Fig.~\ref{fig:two-bp-magnetisation}n).

\subsection{Parameter-space diagram and energy density}
\label{subsec:parameter-space}
The spatially averaged energy density of a \BP[-]{} configuration depends on the
number of Bloch points, their individual types, and the strip geometry.
Furthermore, different spatial arrangements can be realised, \emph{e.g.} four
\BP{}s on a line, or in the corners of a rectangle or diamond shape. Here, we
only consider magnetisation configurations containing between one and eight
\BP{}s in a row, distributed in $x$ direction (strip length) and centred in
$y$ direction (strip width).

In Fig.~\ref{fig:two-bp-magnetisation} we have seen the two fundamentally
different configurations containing two \BP{}s, head-to-head and tail-to-tail
(HH-TT), and head-to-head and head-to-head (HH-HH). Now we investigate a system
containing three \BP{}s. In total, eight configurations can be realised. Three
configurations are fundamentally different, namely HH-HH-HH, HH-HH-TT, and
HH-TT-HH, because they contain distinct numbers of additional antivortices. The
other five configurations are equivalent either because HH and TT swap roles
(\emph{e.g.} TT-TT-TT), or because of the symmetry of the system geometry along
the $x$ axis (\emph{e.g.} TT-HH-HH).

\begin{figure}
  \includegraphics[width=\linewidth]{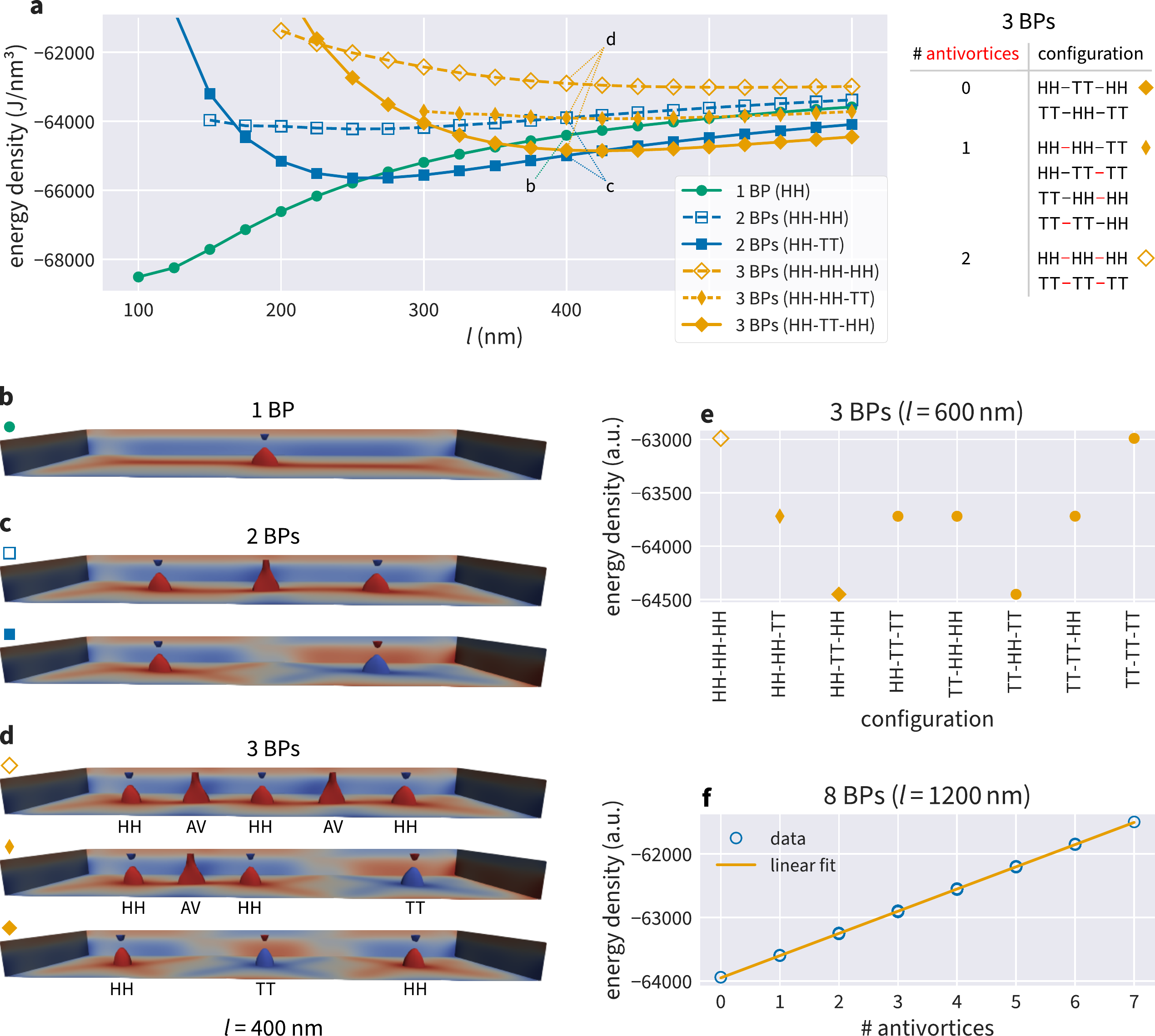}
  \caption{\label{fig:energy-density}(a) Energy densities for energetically
    different \BP[-]{} configurations containing at most three \BP{}s for
    different strip lengths $l$ at a strip width $w=100\,\mathrm{nm}$.
    Simulations have been performed in steps of $\delta l = 25\,\mathrm{nm}$,
    the solid lines are shown to guide the eye. (b -- d) Magnetisation profiles
    for the six different configurations shown in (a) at $l=400\,\mathrm{nm}$.
    Isosurfaces show $m_{z}=\pm 0.9$, colour indicates the z-component. (e)
    Energy density for all possible configurations containing three Bloch
    points. The first three configurations are show in (d) as indicated with the
    distinct marker symbols (at a different strip length). The table in (a)
    lists all different configuration containing three \BP{}s, highlighting the
    number and position of the additional antivortices in the different
    configurations. (f) Energy densities for all configurations containing eight
    \BP{}s. Energy densities for a fixed number of antivortices are nearly
    identical but cause some ``smearing'' of the marker symbols.}
\end{figure}

The fundamentally different configurations (HH-HH-HH, HH-HH-TT, HH-TT-HH) are
shown in Fig.~\ref{fig:energy-density}d in a system with strip length
$l=400\,\mathrm{nm}$. We find one and two additional antivortices (AVs) for
configurations HH-HH-TT and HH-HH-HH, respectively. The table in
Fig.~\ref{fig:energy-density} lists all eight configurations and the respective
number of antivortices.

For each of the three fundamentally different configurations, we compute the
spatially averaged energy density ($E/V$) and plot the three values in
Fig.~\ref{fig:energy-density}a (at $l=400\,\mathrm{nm}$). We find that the
HH-TT-HH configuration has the lowest energy density and the HH-HH-HH
configuration the highest energy density. We will later discuss in more detail
that the presence of antivortices between \BP{}s generally increases the energy
density of the system. The alternating configuration HH-TT-HH does not contain
any antivortices, as these are only needed to mediate between neighbouring
\BP{}s of the same type.

The three yellow lines (filled and open diamonds) in
Fig.~\ref{fig:energy-density}a show the spatially averaged energy density for
the three different configurations as a function of strip length. Not all
configurations are stable for all strip lengths: for example the HH-HH-TT
configuration is only stable for $l\ge 300\,\mathrm{nm}$. If we try to create
the three-\BP[-]{} configuration HH-HH-TT in a shorter nanostrip, \emph{e.g.} at
$l=275\,\mathrm{nm}$, then the configuration is not stable and will change into
a lower-energy configuration, in this case the HH-TT configuration containing
only two \BP{}s. We can see that the energy generally increases with increasing
number of antivortices as mentioned in the previous paragraph. However, there is
a deviation visible for $l\leq 225\,\mathrm{nm}$ where the HH-HH-HH
configuration has a lower energy density than the HH-TT-HH configuration. This
deviation is a result of the short strip length near the stability limit. We
exclude these regions near the minimal stability strip length in the rest of our
discussion.

The blue filled and open squares in Fig.~\ref{fig:energy-density}a show the
energy density for a system containing only two \BP{}s. The corresponding
magnetisation field for $l=400\,\mathrm{nm}$ is visualised in
Fig.~\ref{fig:energy-density}c, and in more detail in
Fig.~\ref{fig:two-bp-magnetisation}. The green circles in
Fig.~\ref{fig:energy-density}a show the energy density for a single \BP{}, and
its magnetisation configuration for $l=400\,\mathrm{nm}$ is shown in
Fig.~\ref{fig:energy-density}b.

For a given strip length $l$ we describe the configuration with the lowest
energy density as the energetically most favourable configuration: below
$l=250\,\nm$ a single \BP{} (green circles) has the lowest energy density. Two
opposite-type \BP{}s (blue squares) have the lowest energy density for
$250\,\nm< l <=400\,\nm$ and three \BP{}s of alternating opposite type (yellow
diamonds) have the lower energy density above $l=400\,\nm$.

Figure~\ref{fig:energy-density}e shows a representation of the energy densities
for all possible three-\BP[-]{} configurations at $l=600\,\mathrm{nm}$. As
already discussed, there are three fundamentally different configurations
characterised by the number of additional antivortices contained in the
configuration (as show in the table in Fig.~\ref{fig:energy-density}). Different
realisations of the same configuration type (swapping HH and TT or using the
strip symmetry) exhibit the same energy density.

\begin{figure}
  \centering
  \includegraphics[width=\linewidth]{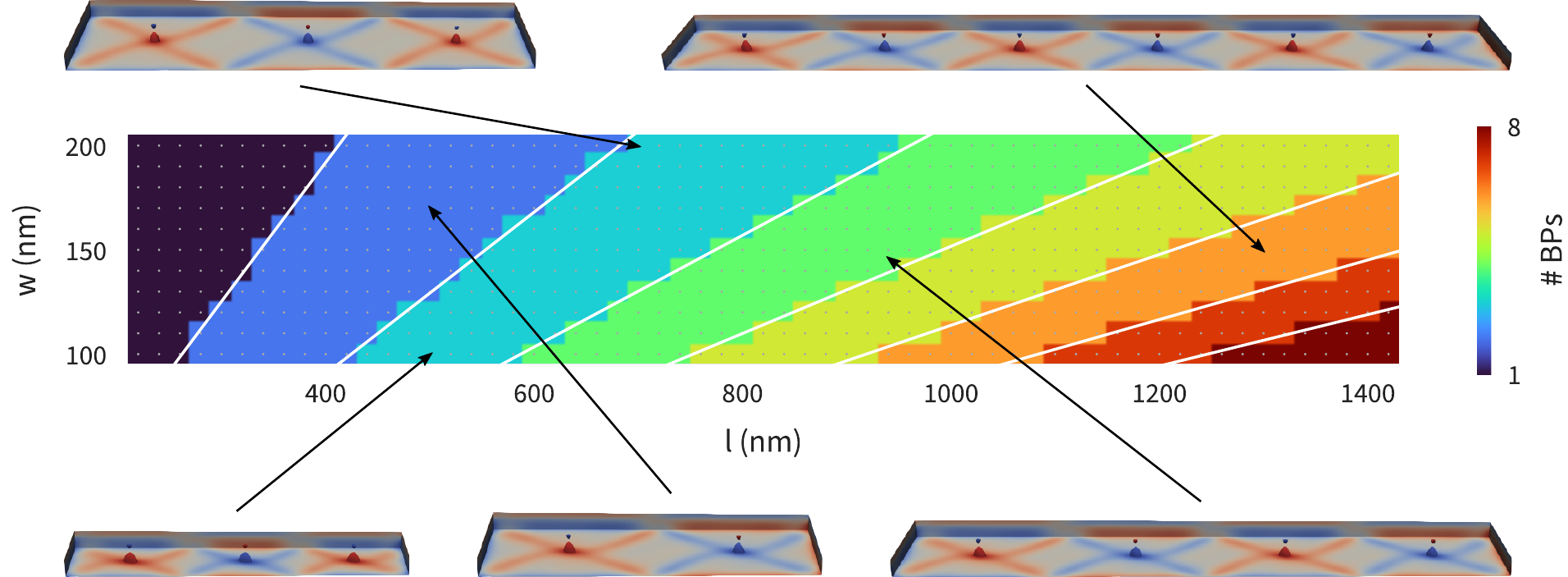}
  \caption{\label{fig:parameter-space-diagram}Parameter-space diagram showing
    the energetically most favourable \BP{} number as a function of length $l$
    and width $w$. All configurations contain \BP{}s of alternating opposite
    type. Magnetisation profiles for selected configurations reveal the
    similarity of the different configurations, isosurfaces show $m_{z}=\pm
    0.9$.}
\end{figure}

In Fig.~\ref{fig:energy-density}a we have seen that the number of \BP{}s in the
energetically most favourable configuration changes depending on the strip
length $l$. Figure~\ref{fig:parameter-space-diagram} contains a parameter-space
diagram showing the energetically most favourable configuration as a function of
the strip length and strip width, using the \BP{} number as a label. To create
Fig.~\ref{fig:parameter-space-diagram}, we ask for each strip length $l$ and a
given strip width $w$, which configuration has the lowest energy density. For
example: all data points in Fig.~\ref{fig:energy-density}a are for a width of
$w=100\,\mathrm{nm}$. Close to $l \approx 400\,\mathrm{nm}$, we see that for $l
\le 400\,\mathrm{nm}$ the two-\BP[-]{} configuration HH-TT (blue squares) has
the lowest energy density but that for $l>400\,\mathrm{nm}$ the three-\BP[-]{}
configuration HH-TT-HH (yellow diamonds) has the lowest value.
Figure~\ref{fig:parameter-space-diagram} shows (for $w=100\,\mathrm{nm}$ on the
$y$ axis) that the two-\BP[-]{} configuration has the lowest energy density up to
$l \approx 400\,\mathrm{nm}$, and the three-\BP[-]{} configuration for larger
$l$ (up to $l \approx 600\,\mathrm{nm}$). All configurations with lowest energy
density are of the alternating \BP[-]{} type, \emph{i.e.} left and right
neighbours of a HH \BP{} are always of type TT, and vice versa (see discussion
below).

Figure~\ref{fig:parameter-space-diagram} shows that with increasing strip
length, the number of \BP{}s that are present in the lowest-energy-density
configuration (as shown Fig.~\ref{fig:energy-density}a for $l \le 600\nm$)
increases: for nanostrips with lengths above $l \approx 1300\nm$ and width
$w=100\nm$, we find eight \BP{}s. Furthermore,
Fig.~\ref{fig:parameter-space-diagram} shows that increasing the width of the
nanostrip leads to a reduced number of Bloch points in the energetically most
favourable configuration.

Figure~\ref{fig:parameter-space-diagram} also shows magnetisation profiles for
selected configurations revealing the similarities in the magnetisation profile
in different strip geometries. The isosurfaces show $m_{z} = \pm 0.9$,
colour indicates the $z$ component. All configurations shown in
Fig.~\ref{fig:parameter-space-diagram} contain \BP{}s of alternating opposite
type, \emph{i.e.} all the lowest-energy configurations do not contain
antivortices.

In the discussion of the fundamentally different configurations containing three
\BP{}s, we have noted that the different configurations can be characterised by
the number of additional antivortices contained in the structure.
Figure~\ref{fig:energy-density}f summarises similar findings for eight Bloch
points where configurations can contain between zero antivortices~(\BP{}s of
alternating opposite type) and seven antivortices~(all \BP{}s of the same type).
In total, 256 configurations can be realised. The plot in
Fig.~\ref{fig:energy-density}f shows the data for all configurations and a
linear fit to the data. Different realisations with the same number of
antivortices cannot be distinguished in this plot as their energies are nearly
identical but cause broadening of the marker symbols for intermediate numbers of
antivortices. The energy density increases linearly with the number of
antivortices.

There is an important difference between three \BP{}s
(Fig.~\ref{fig:energy-density}e) and eight \BP{}s
(Fig.~\ref{fig:energy-density}f). For a fixed number of antivortices, all
different three-\BP[-]{} configurations are equivalent because of the system's
symmetry and therefore must have the same energy density. However, for eight
\BP{}s we additionally find that different configurations that are not related
via symmetry also have the same energy density if they contain the same number
of antivortices. For example, the configurations HH-HH-HH-HH-HH-HH-HH-TT and
HH-HH-HH-HH-TT-TT-TT-TT both contain 6 antivortices (located between
neighbouring same-type \BP{}s) but cannot be transformed into each other using
the symmetries we discussed. Yet, they exhibit the same spatially averaged
energy density. Our findings suggest that the energy density of the \BP{}s is
independent of the configuration of other \BP{}s in the system. The energy
density can be obtained from a configuration containing \BP{}s of alternating
opposite type with additional contributions originating from the additional
antivortices between neighbouring same-type Bloch points.

This is the reason for the lowest-energy-density configurations shown in
Fig.~\ref{fig:parameter-space-diagram} consisting of pairs of \BP{}s of
alternating type: for same-type neighbours an antivortex is required to mediate
the magnetisation between the \BP{}s of the same type, and the presence of such
antivortices would increase the spatially averaged energy density.

We can make one additional observation in Fig.~\ref{fig:energy-density}a. The
energy density of a configuration changes as a function of strip length $l$. All
configurations containing two or three \BP{}s have one energy minimum at a
certain length that we call the \emph{optimal length} $l_{\mathrm{o}}$. For
example, the optimal length for the HH-TT configuration (blue filled squares in
Fig.~\ref{fig:energy-density}a) is $l_{\mathrm{o}} \approx 275\,\mathrm{nm}$.

\subsection{Predicting strip geometries for larger systems}
\label{subsec:80-bp}

So far, we have focused on small systems containing at most eight Bloch points.
Based on this information we can predict strip geometries suitable for an
arbitrary number of \BP{}s.

Figure~\ref{fig:energy-density}a shows that meta-stable configurations
containing multiple \BP{}s can be realised over a broad range of strip lengths
but need a certain minimal strip length. This minimal length depends on the
number of \BP{}s and additional antivortices in the configuration. Furthermore,
Fig.~\ref{fig:energy-density}a shows that all configurations have a minimum in
the energy density at a certain optimal length $l_{\mathrm{o}}$.

\begin{figure}
  \includegraphics[width=\linewidth]{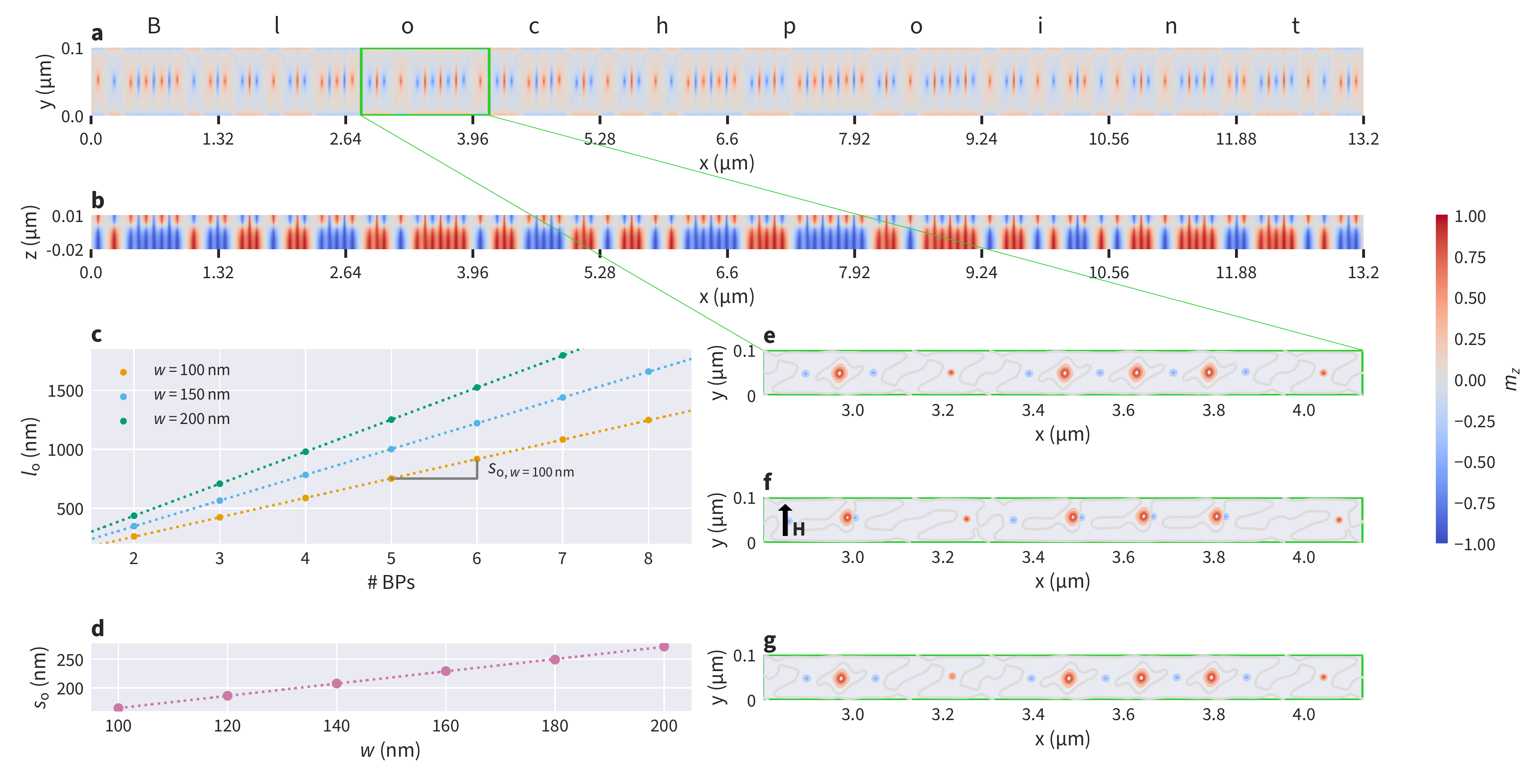}
  \caption{\label{fig:blochpoint-ascii}(a, b) ASCII encoding of the string
    \texttt{Blochpoint} using 80 \BP{}s. Cross sections show (a) the $xy$ plane
    at $z=1\,\mathrm{nm}$ and (b) the $xz$ plane at $y=50\,\mathrm{nm}$. The
    strip length is chosen according to the predicted value for a nanostrip with
    width $w=100\,\mathrm{nm}$. Labels on the $x$ axis mark blocks of eight
    \BP{}s, \emph{i.e.} individual bytes. Note that the aspect ratio is not
    correct in order to improve visibility. (c) Optimal lengths $l_{\mathrm{o}}$
    for two to eight \BP{}s. The fit is used to predict lengths for more than
    eight \BP{}s. (d) Optimal \BP{} distance $s_{\mathrm{o}}$ as a function of
    strip width $w$. (e~--~g) show an enlarged part of the nanostrip (correct
    aspect ratio) as highlighted in (a) to demonstrate the stability of the
    configuration: (e) initial configuration after energy minimisation; (f) an
    external magnetic field $H= 0.25\,\mathrm{T} / \mu_0$ is applied in the $+y$
    direction for $0.5\,\mathrm{ns}$; (g) after removing the external field the
    system evolves freely and converges back to the initial state (snapshot
    after $5\,\mathrm{ns}$). (e~--~g) show contour lines of the $m_z$ component
    to improve visibility of the disturbance introduced by the external magnetic
    field. \BP{}s are located at the small red and blue dots, the larger red
    circles show the additional antivortices.}
\end{figure}

To predict strip geometries suitable for an arbitrary number of \BP{}s we focus
on configurations containing up to eight \BP{}s of alternating opposite type. We
find that the optimal length $l_{\mathrm{o}}$ increases linearly with the number
of \BP{}s (Fig.~\ref{fig:blochpoint-ascii}c) with the slope defining the optimal
\BP{} spacing $s_{\mathrm{o}}$. Furthermore, we find that $s_{\mathrm{o}}$
increases linearly with increasing strip width
(Fig.~\ref{fig:blochpoint-ascii}d). We obtain $s_{\mathrm{o},
  w=100\,\mathrm{nm}} \approx 165\,\mathrm{nm}$ and $s_{\mathrm{o},
  w=200\,\mathrm{nm}} \approx 272\,\mathrm{nm}$ with an estimated accuracy of
$\delta s_{\mathrm{o}} \approx 3\,\mathrm{nm}$. These observations lead to our
working hypothesis that the ideal \BP[-]{} spacing $s_{\mathrm{o}}$ is
independent of the number of \BP{}s in the system and suitable to predict
geometries for more than eight \BP{}s. This prediction can be used for arbitrary
configurations not only alternating opposite-type \BP{}s.

As an illustrative example, we simulate one specific configuration containing 80
\BP{}s, encoding the 10-character word \texttt{Blochpoint} in ASCII code (eight
bits per letter). We simulate a strip with the predicted length $l = 80
s_{\mathrm{o}} = \SI{13.2}{\micro\metre}$ at a width of $w=100\,\mathrm{nm}$ and
with $s_{\mathrm{o}} = 165\,\mathrm{nm}$.

We minimise the energy of a suitable initial configuration resulting in the
80-bit configuration as shown in Fig.~\ref{fig:blochpoint-ascii}a,~b. The cross
sections show the $xy$ plane at $z=1\,\mathrm{nm}$
(Fig.~\ref{fig:blochpoint-ascii}a) and the $xz$ plane at $y=50\,\mathrm{nm}$
(Fig.~\ref{fig:blochpoint-ascii}b). Note that the aspect ratio is not correct in
order to improve visibility. Figure~\ref{fig:blochpoint-ascii}e shows contour
lines for $m_z$ for a part of the nanostrip (correct aspect ratio) as indicated
in Fig.~\ref{fig:blochpoint-ascii}a. \BP{}s in Fig.~\ref{fig:blochpoint-ascii}e
are located at the small red and
blue dots. The larger red circles ($m_z > 0.5$) show antivortices between
same-type \BP{}s.

To test the stability of the 80-\BP[-]{} configuration we apply a short magnetic
field pulse in the $+y$ direction ($H=25\,\mathrm{mT}/\mu_0$, applied for
$t=0.5\,\mathrm{ns}$). The modified magnetisation field configuration at the end
of the $0.5\,\mathrm{ns}$ period is shown in Fig.~\ref{fig:blochpoint-ascii}f.
Then, we set the applied field back to zero, and let the system evolve freely by
carrying out a time-integration. We find that the magnetisation converges back
to the initial state: Fig.~\ref{fig:blochpoint-ascii}g shows the configuration
after $t=5\,\mathrm{ns}$ of free relaxation.

To understand the robustness of the predicted geometry, we vary the strip length
$l$ and find that the desired 80-\BP[-]{} configuration can be stabilised over a
range of strip geometries. The minimal strip length is around $0.66
l_{\mathrm{o}}$ the maximal strip length around $4l_{\mathrm{o}}$.

Within the range of stability of the 80-bit configuration ($0.66l_\mathrm{o} \le
l \le 4l_\mathrm{o}$), we find that the length $l_\mathrm{o}$ is closer to the
lower stability boundary ($\approx 0.66 l_\mathrm{o}$) than to the upper limit
($\approx 4l_\mathrm{o}$). This is consistent with the energy density curve for
the HH-TT configuration in Fig.~\ref{fig:energy-density}a (blue filled squares)
where we see that the energy density as a function of the strip length is
asymmetric, and that its energy minimum, located at strip length
$l_{\mathrm{o}}$, is located at a comparatively small strip length within the
range of possible strip lengths over which the configuration is meta-stable
(stability limits are not visible in Fig.~\ref{fig:energy-density}).

\section{Discussion}
\label{sec:discussion}

The \BP[-]{} configuration originates from vortices with identical circularity,
but opposite polarisation, which are stabilised through the DMI of the material,
which fixes the core orientation relative to circularity through the left- or
right-handed chirality. The \BP{}s form an interesting topological excitation in
a helimagnetic system, which extends the set of well-known magnetic structures
such as domain walls, vortices, and skyrmions. In the geometry described here,
the \BP{}s are in equilibrium and can be manipulated.

We have found remarkable features of multiple interacting \BP{}s in two-layer
nanostrips. The two different types -- head-to-head (HH) and tail-to-tail (TT)
-- can be geometrically arranged in any arbitrary order, and this magnetisation
configuration resembles a meta-stable configuration (within certain constraints
on the strip width and length). The spatially averaged energy density for a
system with $n$ \BP{}s increases in fixed steps. The number of steps scales --
within our accuracy -- exactly linearly with the number of antivortices in the
configuration (or equivalently: the number of neighbouring same-type \BP{}s). We
can determine an optimal \BP[-]{} spacing $s_{\mathrm{o}}$ between \BP{}s within
a line of \BP{}s (corresponding to a distance over which a \BP{} extends).

In the following, we speculate about possible future applications of \BP{}s. One
key-feature distinguishing \BP{}s from many other particle-like magnetic
configurations is the demonstrated coexistence of \BP{}s of two different types
in a single sample making \BP{}s an interesting candidate for binary data
representation. In the racetrack-like designs~\cite{Parkin2008, Sampaio2013},
when realised with magnetic excitations of which only one type exists -- such as
skyrmions -- we need to ensure that skyrmions keep their relative positions to
be able to interpret the presence of a skyrmion as 1 and the absence of a
skyrmion as 0. The two different types of \BP{}s presented here could be used to
encode binary data without the need to rely on fixed spacing of magnetic
objects: a HH configuration could represent ``\texttt{1}'' and a TT
configuration could represent ``\texttt{0}''. In the context of skyrmion-based
realisation of the racetrack approach, other ideas to overcome the fixed-spacing
requirement include the use of a combination of skyrmion tubes and chiral
bobbers~\cite{Zheng2018} and the two-lane racetrack memory~\cite{Muller2017}.

On the way towards possible applications of \BP{}s many more questions need to
be addressed. These are related to \BP{} manipulation (movement, switching,
creation/annihilation), sensing of \BP{}s, and to the thermal stability of
\BP{}s in general and energy barriers between different configurations
containing multiple \BP{}s.

In summary, we have demonstrated that two-layer FeGe nanostrips can host
multiple \BP{}s in any combination of head-to-head and tail-to-tail. Based on
our simulations containing up to eight \BP{}s, we can predict strip geometries
suitable for an arbitrary number of \BP{}s. We have verified this prediction by
studying a system containing 80 \BP{}s.

All results obtained in this work can be reproduced from the repository in
Ref.~\cite{Lang2022} which contains Jupyter notebooks~\cite{Granger2021} to
rerun the micromagnetic simulations and recreate all plots. In the repository
pre-computed datasets are also available.
\section{Methods}
\subsection{System}
We simulate rectangular two-layer nanostrips with opposite chirality (opposite
sign of $D$) in the two layers. We vary strip length and width, the thickness of
both layers is fixed (bottom layer: $20\nm$, top layer: $10\nm$). We focus on up
to eight \BP{}s and accordingly choose nanostrips with lengths between
$100\nm$ and $1400\nm$, and widths between $100\nm$ and $200\nm$. The energy
equation
\begin{equation}
  E = \int \mathrm{d}^3r \left( w_{\mathrm{ex}} + w_{\mathrm{dmi}} + w_{\mathrm{d}} \right)
\end{equation}
contains exchange energy density $w_{\mathrm{ex}}$, bulk Dzyaloshinskii-Moriya
energy density $w_{\mathrm{dmi}}$, and demagnetisation energy density
$w_{\mathrm{d}}$. The magnetisation dynamics is simulated using the
Landau-Lifshitz-Gilbert equation~\cite{Landau1935, Gilbert2004}:
\begin{equation}
  \frac{\partial\mathbf{m}}{\partial t} = \gamma^* \mathbf{m} \times \mathbf{H}_{\mathrm{eff}} + \alpha \mathbf{m} \times \frac{\partial \mathbf{m}}{\partial t},
\end{equation}
where $\gamma^* = \gamma (1 + \alpha^2)$, with $\gamma$ being the gyromagnetic
ratio and $\alpha$ Gilbert damping. Material parameters are based on
FeGe~\cite{Beg2015}: $A = 8.87\,\mathrm{pJ}\,\mathrm{m}^{-1}$, $D =
1.58\,\mathrm{mJ}\,\mathrm{m}^{-2}$, $M_{\mathrm{s}} =
384\,\mathrm{kA}\,\mathrm{m}^{-1}$, $\alpha=0.28$. We use finite-difference
micromagnetic simulations to minimise the energy. All simulations are done using
Ubermag~\cite{Beg2022, Beg2017a, Beg2022a} with OOMMF~\cite{Donahue1999} as
computational backend and an extension for DMI of
crystalclass~T~\cite{Cortes-Ortuno2018, Cortes-Ortuno2018a}.

\subsection{Simulation procedure}
All simulations in this study follow a three-step initialisation and
minimisation scheme: (i)~initialisation, (ii)~fixed minimisation, (iii)~free
minimisation. In the micromagnetic framework the system is studied at zero
temperature, \emph{i.e.} without thermal fluctuations. Therefore, it is only
possible to find local minima that are accessible from the initial
configuration. Starting from experimentally feasible initial configurations,
such as full saturation, we are able to find magnetisation configurations
containing a single or multiple Bloch points depending on the strip geometry.

To facilitate the process of studying arbitrary \BP[-]{} configurations,
independent of the strip geometry in a systematic way, we have developed a
simulation scheme that guarantees a magnetisation configuration containing a
predictable number of \BP{}s. We note that this scheme can probably not be
applied directly to an experimental set-up.

For the initialisation, step~(i), we start by dividing the nanostrip into
equally sized regions (in $x$ direction), one region per Bloch point. To enforce
the formation of a Bloch point, the magnetisation in each region is initialised
as follows: for a head-to-head \BP{} we initialise the centre region of the
topmost layer of cells with $\mathbf{m} = (0, 0, -1)$ and all other cells with
$\mathbf{m} = (0, 0, 1)$. A region hosting a tail-to-tail Bloch point is
initialised with reversed $z$ component of the magnetisation (see supplementary
Fig.~1 for a schematic plot of the different subregions). We then minimise the
energy in two steps (supplementary Fig.~2). During the first energy
minimisation, step (ii), we keep the magnetisation of the topmost cells --
initialised with reversed magnetisation -- and a similarly-sized layer of cells
at the bottom sample boundary fixed. This ensures the formation of a \BP{} at
the interface between the two layers. The second energy minimisation, step
(iii), is done without any fixed cells, \emph{i.e.} magnetisation in all cells
can freely change, and \BP{}s could move in any direction to further minimise
the energy of the configuration. In this step, the system can find the local
energy minimum.

\subsection{Classification}
In the micromagnetic framework, it is not possible to directly observe Bloch
points because of the fixed norm of the magnetisation vector. A single Bloch
point is characterised by the integral value of the topological charge density
over a closed surface $A$ surrounding the Bloch point~\cite{Im2019}:
\begin{equation}
  S = \frac{1}{4\pi} \int_A \mathrm{d}\mathbf{A} \cdot \mathbf{F} = \pm 1,
\end{equation}
where $\mathbf{F}$ is the emergent magnetic field~\cite{Volovik1987,
  Liu2018}. The components of $\mathbf{F}$ are defined as:
\begin{equation}
  F_i = \mathbf{m} \cdot \left( \partial_j \mathbf{m} \times \partial_k \mathbf{m} \right),
\end{equation}
where $(i, j, k)$ is an even permutation of $(x, y, z)$. To detect a single
\BP{} in a sample the integral can be computed over the whole sample
surface and the exact position of the \BP{} does not need to be known.

\begin{figure}
  \includegraphics[width=\linewidth]{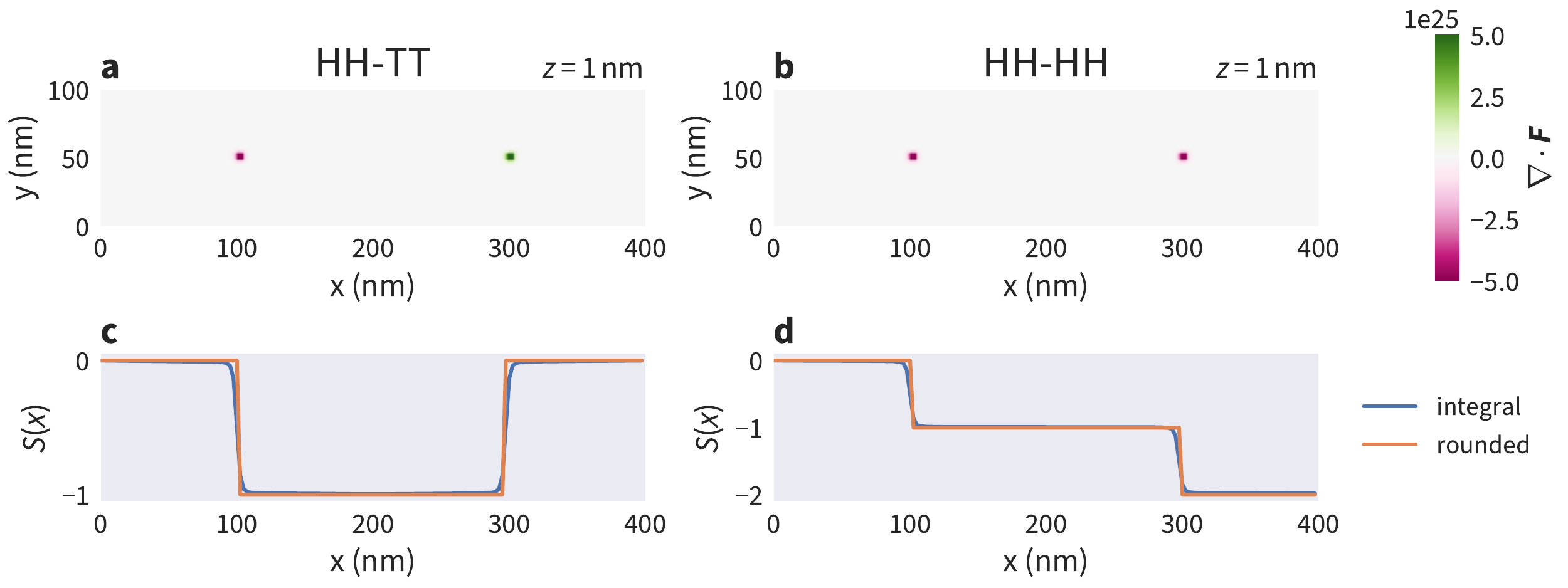}
  \caption{\label{fig:two-bp-classification}Classification of the two
    configurations containing two \BP{}s shown in
    Fig.~\ref{fig:two-bp-magnetisation}. The divergence of the emergent magnetic
    field for the two opposite- and same-type \BP{}s is shown in panels (a) and
    (b), respectively. The $xy$ plane visualised here is located at
    $z=1\,\mathrm{nm}$, just above the interface. (c, d) The result of the
    convolution \eqref{eq:S-of-x} which is used to identify the occurrence of
    \BP{}s and their type due to the steps $\Delta S = \pm 1$.}
\end{figure}

This method is not directly applicable to multiple Bloch points when their
positions are unknown: the sign of the topological charge of a \BP{} depends on
its type (HH: $S=-1$, TT: $S=+1$). Therefore, contributions to the surface
integral from \BP{}s of opposite type cancel out.
Figure~\ref{fig:two-bp-classification} shows the divergence of the emergent
field $\nabla\cdot\mathbf{F}$ for a HH and a TT \BP{} (a) and two HH \BP{}s (b),
the two configurations discussed in Fig.~\ref{fig:two-bp-magnetisation}.

To classify nanostrips that potentially contain multiple \BP{}s we compute the
convolution of the divergence of the emergent magnetic field with a Heaviside
step function $\Theta$:
\begin{equation}
S(x) = \frac{1}{4\pi} \int_{V'}\mathrm{d}^3r' \, \Theta(x - x') \nabla_{\mathbf{r}'} \cdot \mathbf{F}(\mathbf{r}'). \label{eq:S-of-x}
\end{equation}
Due to numerical inaccuracies the result of the integral deviates from integer
values. By translating the surface integral into a volume integral over the
divergence of the emergent magnetic field using the divergence theorem the
accuracy can be improved by roughly one order of magnitude.

In our set-up \BP{}s are expected to be distributed along $x$ following the
strip geometry which justifies computing $S$ as a function of $x$. This
convolution can be interpreted as computing a series of integrals over
increasing subvolumes $V'$ of the nanostrip starting at the left boundary
($x=0\,\mathrm{nm}$). We round $S(x)$ to integer values and count steps $\Delta
S$ in this function.

Figure~\ref{fig:two-bp-classification}c and~d show $S(x)$ for the two example
configurations. A head-to-head \BP{} is identified by $\Delta S = -1$, a
tail-to-tail \BP{} by $\Delta S = +1$ corresponding to the topological charge of
a \BP{} being $S = \pm 1$. Rounding to integer values is justified because
deviations from integer values in the integral are a direct consequence of the
limited accuracy due to the discretisation. The deviation from integer values
decreases with decreasing cell size (see supplementary Fig.~3 for details).

\bibliography{references}

\begin{thebibliography}{24}%
\makeatletter
\providecommand \@ifxundefined [1]{%
 \@ifx{#1\undefined}
}%
\providecommand \@ifnum [1]{%
 \ifnum #1\expandafter \@firstoftwo
 \else \expandafter \@secondoftwo
 \fi
}%
\providecommand \@ifx [1]{%
 \ifx #1\expandafter \@firstoftwo
 \else \expandafter \@secondoftwo
 \fi
}%
\providecommand \natexlab [1]{#1}%
\providecommand \enquote  [1]{``#1''}%
\providecommand \bibnamefont  [1]{#1}%
\providecommand \bibfnamefont [1]{#1}%
\providecommand \citenamefont [1]{#1}%
\providecommand \href@noop [0]{\@secondoftwo}%
\providecommand \href [0]{\begingroup \@sanitize@url \@href}%
\providecommand \@href[1]{\@@startlink{#1}\@@href}%
\providecommand \@@href[1]{\endgroup#1\@@endlink}%
\providecommand \@sanitize@url [0]{\catcode `\\12\catcode `\$12\catcode
  `\&12\catcode `\#12\catcode `\^12\catcode `\_12\catcode `\%12\relax}%
\providecommand \@@startlink[1]{}%
\providecommand \@@endlink[0]{}%
\providecommand \url  [0]{\begingroup\@sanitize@url \@url }%
\providecommand \@url [1]{\endgroup\@href {#1}{\urlprefix }}%
\providecommand \urlprefix  [0]{URL }%
\providecommand \Eprint [0]{\href }%
\providecommand \doibase [0]{https://doi.org/}%
\providecommand \selectlanguage [0]{\@gobble}%
\providecommand \bibinfo  [0]{\@secondoftwo}%
\providecommand \bibfield  [0]{\@secondoftwo}%
\providecommand \translation [1]{[#1]}%
\providecommand \BibitemOpen [0]{}%
\providecommand \bibitemStop [0]{}%
\providecommand \bibitemNoStop [0]{.\EOS\space}%
\providecommand \EOS [0]{\spacefactor3000\relax}%
\providecommand \BibitemShut  [1]{\csname bibitem#1\endcsname}%
\let\auto@bib@innerbib\@empty
\bibitem [{\citenamefont {Wang}\ and\ \citenamefont {Wang}(2021)}]{Wang2021}%
  \BibitemOpen
  \bibfield  {author} {\bibinfo {author} {\bibfnamefont {X.~S.}\ \bibnamefont
  {Wang}}\ and\ \bibinfo {author} {\bibfnamefont {X.~R.}\ \bibnamefont
  {Wang}},\ }\bibfield  {title} {\bibinfo {title} {Topology in {{Magnetism}}},\
  }in\ \href {https://doi.org/10.1007/978-3-030-62844-4_14} {\emph {\bibinfo
  {booktitle} {Chirality, {{Magnetism}} and {{Magnetoelectricity}}: {{Separate
  Phenomena}} and {{Joint Effects}} in {{Metamaterial Structures}}}}},\
  \bibinfo {series and number} {Topics in {{Applied Physics}}},\ \bibinfo
  {editor} {edited by\ \bibinfo {editor} {\bibfnamefont {E.}~\bibnamefont
  {Kamenetskii}}}\ (\bibinfo  {publisher} {{Springer International
  Publishing}},\ \bibinfo {address} {{Cham}},\ \bibinfo {year} {2021})\ pp.\
  \bibinfo {pages} {357--403}\BibitemShut {NoStop}%
\bibitem [{\citenamefont {G{\"o}bel}\ \emph {et~al.}(2021)\citenamefont
  {G{\"o}bel}, \citenamefont {Mertig},\ and\ \citenamefont
  {Tretiakov}}]{Gobel2021}%
  \BibitemOpen
  \bibfield  {author} {\bibinfo {author} {\bibfnamefont {B.}~\bibnamefont
  {G{\"o}bel}}, \bibinfo {author} {\bibfnamefont {I.}~\bibnamefont {Mertig}},\
  and\ \bibinfo {author} {\bibfnamefont {O.~A.}\ \bibnamefont {Tretiakov}},\
  }\bibfield  {title} {\bibinfo {title} {Beyond skyrmions: {{Review}} and
  perspectives of alternative magnetic quasiparticles},\ }\href
  {https://doi.org/10.1016/j.physrep.2020.10.001} {\bibfield  {journal}
  {\bibinfo  {journal} {Physics Reports}\ }\bibinfo {series} {Beyond Skyrmions:
  {{Review}} and Perspectives of Alternative Magnetic Quasiparticles},\ \textbf
  {\bibinfo {volume} {895}},\ \bibinfo {pages} {1} (\bibinfo {year}
  {2021})}\BibitemShut {NoStop}%
\bibitem [{\citenamefont {Feldtkeller}(1965)}]{Feldtkeller1965}%
  \BibitemOpen
  \bibfield  {author} {\bibinfo {author} {\bibfnamefont {E.}~\bibnamefont
  {Feldtkeller}},\ }\bibfield  {title} {\bibinfo {title} {Mikromagnetisch
  stetige und unstetige {{Magnetisierungskonfigurationen}}},\ }\href@noop {}
  {\bibfield  {journal} {\bibinfo  {journal} {Zeitschrift fur Angewandte
  Physik}\ }\textbf {\bibinfo {volume} {19}},\ \bibinfo {pages} {530} (\bibinfo
  {year} {1965})}\BibitemShut {NoStop}%
\bibitem [{\citenamefont {D{\"o}ring}(1968)}]{Doring1968}%
  \BibitemOpen
  \bibfield  {author} {\bibinfo {author} {\bibfnamefont {W.}~\bibnamefont
  {D{\"o}ring}},\ }\bibfield  {title} {\bibinfo {title} {Point
  {{Singularities}} in {{Micromagnetism}}},\ }\href
  {https://doi.org/10.1063/1.1656144} {\bibfield  {journal} {\bibinfo
  {journal} {Journal of Applied Physics}\ }\textbf {\bibinfo {volume} {39}},\
  \bibinfo {pages} {1006} (\bibinfo {year} {1968})}\BibitemShut {NoStop}%
\bibitem [{\citenamefont {Beg}\ \emph {et~al.}(2019)\citenamefont {Beg},
  \citenamefont {Pepper}, \citenamefont {{Cort{\'e}s-Ortu{\~n}o}},
  \citenamefont {Atie}, \citenamefont {Bisotti}, \citenamefont {Downing},
  \citenamefont {Kluyver}, \citenamefont {Hovorka},\ and\ \citenamefont
  {Fangohr}}]{Beg2019}%
  \BibitemOpen
  \bibfield  {author} {\bibinfo {author} {\bibfnamefont {M.}~\bibnamefont
  {Beg}}, \bibinfo {author} {\bibfnamefont {R.~A.}\ \bibnamefont {Pepper}},
  \bibinfo {author} {\bibfnamefont {D.}~\bibnamefont
  {{Cort{\'e}s-Ortu{\~n}o}}}, \bibinfo {author} {\bibfnamefont
  {B.}~\bibnamefont {Atie}}, \bibinfo {author} {\bibfnamefont {M.-A.}\
  \bibnamefont {Bisotti}}, \bibinfo {author} {\bibfnamefont {G.}~\bibnamefont
  {Downing}}, \bibinfo {author} {\bibfnamefont {T.}~\bibnamefont {Kluyver}},
  \bibinfo {author} {\bibfnamefont {O.}~\bibnamefont {Hovorka}},\ and\ \bibinfo
  {author} {\bibfnamefont {H.}~\bibnamefont {Fangohr}},\ }\bibfield  {title}
  {\bibinfo {title} {Stable and manipulable {{Bloch}} point},\ }\href
  {https://doi.org/10.1038/s41598-019-44462-2} {\bibfield  {journal} {\bibinfo
  {journal} {Scientific Reports}\ }\textbf {\bibinfo {volume} {9}},\ \bibinfo
  {pages} {7959} (\bibinfo {year} {2019})}\BibitemShut {NoStop}%
\bibitem [{\citenamefont {Behncke}\ \emph {et~al.}(2018)\citenamefont
  {Behncke}, \citenamefont {Adolff},\ and\ \citenamefont
  {Meier}}]{Behncke2018}%
  \BibitemOpen
  \bibfield  {author} {\bibinfo {author} {\bibfnamefont {C.}~\bibnamefont
  {Behncke}}, \bibinfo {author} {\bibfnamefont {C.~F.}\ \bibnamefont
  {Adolff}},\ and\ \bibinfo {author} {\bibfnamefont {G.}~\bibnamefont
  {Meier}},\ }\bibfield  {title} {\bibinfo {title} {Magnetic {{Vortices}}},\
  }in\ \href {https://doi.org/10.1007/978-3-319-97334-0_3} {\emph {\bibinfo
  {booktitle} {Topology in {{Magnetism}}}}},\ \bibinfo {series and number}
  {Springer {{Series}} in {{Solid-State Sciences}}},\ \bibinfo {editor} {edited
  by\ \bibinfo {editor} {\bibfnamefont {J.}~\bibnamefont {Zang}}, \bibinfo
  {editor} {\bibfnamefont {V.}~\bibnamefont {Cros}},\ and\ \bibinfo {editor}
  {\bibfnamefont {A.}~\bibnamefont {Hoffmann}}}\ (\bibinfo  {publisher}
  {{Springer International Publishing}},\ \bibinfo {address} {{Cham}},\
  \bibinfo {year} {2018})\ pp.\ \bibinfo {pages} {75--115}\BibitemShut
  {NoStop}%
\bibitem [{\citenamefont {Parkin}\ \emph {et~al.}(2008)\citenamefont {Parkin},
  \citenamefont {Hayashi},\ and\ \citenamefont {Thomas}}]{Parkin2008}%
  \BibitemOpen
  \bibfield  {author} {\bibinfo {author} {\bibfnamefont {S.~S.~P.}\
  \bibnamefont {Parkin}}, \bibinfo {author} {\bibfnamefont {M.}~\bibnamefont
  {Hayashi}},\ and\ \bibinfo {author} {\bibfnamefont {L.}~\bibnamefont
  {Thomas}},\ }\bibfield  {title} {\bibinfo {title} {Magnetic {{Domain-Wall
  Racetrack Memory}}},\ }\href {https://doi.org/10.1126/science.1145799}
  {\bibfield  {journal} {\bibinfo  {journal} {Science}\ }\textbf {\bibinfo
  {volume} {320}},\ \bibinfo {pages} {190} (\bibinfo {year}
  {2008})}\BibitemShut {NoStop}%
\bibitem [{\citenamefont {Sampaio}\ \emph {et~al.}(2013)\citenamefont
  {Sampaio}, \citenamefont {Cros}, \citenamefont {Rohart}, \citenamefont
  {Thiaville},\ and\ \citenamefont {Fert}}]{Sampaio2013}%
  \BibitemOpen
  \bibfield  {author} {\bibinfo {author} {\bibfnamefont {J.}~\bibnamefont
  {Sampaio}}, \bibinfo {author} {\bibfnamefont {V.}~\bibnamefont {Cros}},
  \bibinfo {author} {\bibfnamefont {S.}~\bibnamefont {Rohart}}, \bibinfo
  {author} {\bibfnamefont {A.}~\bibnamefont {Thiaville}},\ and\ \bibinfo
  {author} {\bibfnamefont {A.}~\bibnamefont {Fert}},\ }\bibfield  {title}
  {\bibinfo {title} {Nucleation, stability and current-induced motion of
  isolated magnetic skyrmions in nanostructures},\ }\href
  {https://doi.org/10.1038/nnano.2013.210} {\bibfield  {journal} {\bibinfo
  {journal} {Nature Nanotechnology}\ }\textbf {\bibinfo {volume} {8}},\
  \bibinfo {pages} {839} (\bibinfo {year} {2013})}\BibitemShut {NoStop}%
\bibitem [{\citenamefont {Zheng}\ \emph {et~al.}(2018)\citenamefont {Zheng},
  \citenamefont {Rybakov}, \citenamefont {Borisov}, \citenamefont {Song},
  \citenamefont {Wang}, \citenamefont {Li}, \citenamefont {Du}, \citenamefont
  {Kiselev}, \citenamefont {Caron}, \citenamefont {Kov{\'a}cs}, \citenamefont
  {Tian}, \citenamefont {Zhang}, \citenamefont {Bl{\"u}gel},\ and\
  \citenamefont {{Dunin-Borkowski}}}]{Zheng2018}%
  \BibitemOpen
  \bibfield  {author} {\bibinfo {author} {\bibfnamefont {F.}~\bibnamefont
  {Zheng}}, \bibinfo {author} {\bibfnamefont {F.~N.}\ \bibnamefont {Rybakov}},
  \bibinfo {author} {\bibfnamefont {A.~B.}\ \bibnamefont {Borisov}}, \bibinfo
  {author} {\bibfnamefont {D.}~\bibnamefont {Song}}, \bibinfo {author}
  {\bibfnamefont {S.}~\bibnamefont {Wang}}, \bibinfo {author} {\bibfnamefont
  {Z.-A.}\ \bibnamefont {Li}}, \bibinfo {author} {\bibfnamefont
  {H.}~\bibnamefont {Du}}, \bibinfo {author} {\bibfnamefont {N.~S.}\
  \bibnamefont {Kiselev}}, \bibinfo {author} {\bibfnamefont {J.}~\bibnamefont
  {Caron}}, \bibinfo {author} {\bibfnamefont {A.}~\bibnamefont {Kov{\'a}cs}},
  \bibinfo {author} {\bibfnamefont {M.}~\bibnamefont {Tian}}, \bibinfo {author}
  {\bibfnamefont {Y.}~\bibnamefont {Zhang}}, \bibinfo {author} {\bibfnamefont
  {S.}~\bibnamefont {Bl{\"u}gel}},\ and\ \bibinfo {author} {\bibfnamefont
  {R.~E.}\ \bibnamefont {{Dunin-Borkowski}}},\ }\bibfield  {title} {\bibinfo
  {title} {Experimental observation of chiral magnetic bobbers in {{B20-type
  FeGe}}},\ }\href {https://doi.org/10.1038/s41565-018-0093-3} {\bibfield
  {journal} {\bibinfo  {journal} {Nature Nanotechnology}\ }\textbf {\bibinfo
  {volume} {13}},\ \bibinfo {pages} {451} (\bibinfo {year} {2018})}\BibitemShut
  {NoStop}%
\bibitem [{\citenamefont {M{\"u}ller}(2017)}]{Muller2017}%
  \BibitemOpen
  \bibfield  {author} {\bibinfo {author} {\bibfnamefont {J.}~\bibnamefont
  {M{\"u}ller}},\ }\bibfield  {title} {\bibinfo {title} {Magnetic skyrmions on
  a two-lane racetrack},\ }\href {https://doi.org/10.1088/1367-2630/aa5b55}
  {\bibfield  {journal} {\bibinfo  {journal} {New Journal of Physics}\ }\textbf
  {\bibinfo {volume} {19}},\ \bibinfo {pages} {025002} (\bibinfo {year}
  {2017})}\BibitemShut {NoStop}%
\bibitem [{\citenamefont {Lang}\ \emph {et~al.}(2022)\citenamefont {Lang},
  \citenamefont {Beg}, \citenamefont {Hovorka},\ and\ \citenamefont
  {Fangohr}}]{Lang2022}%
  \BibitemOpen
  \bibfield  {author} {\bibinfo {author} {\bibfnamefont {M.}~\bibnamefont
  {Lang}}, \bibinfo {author} {\bibfnamefont {M.}~\bibnamefont {Beg}}, \bibinfo
  {author} {\bibfnamefont {O.}~\bibnamefont {Hovorka}},\ and\ \bibinfo {author}
  {\bibfnamefont {H.}~\bibnamefont {Fangohr}},\ }\href
  {https://doi.org/10.5281/zenodo.6384937} {\bibinfo {title} {Bloch points in
  nanostrips}},\ \bibinfo {howpublished} {Zenodo} (\bibinfo {year}
  {2022})\BibitemShut {NoStop}%
\bibitem [{\citenamefont {Granger}\ and\ \citenamefont
  {P{\'e}rez}(2021)}]{Granger2021}%
  \BibitemOpen
  \bibfield  {author} {\bibinfo {author} {\bibfnamefont {B.~E.}\ \bibnamefont
  {Granger}}\ and\ \bibinfo {author} {\bibfnamefont {F.}~\bibnamefont
  {P{\'e}rez}},\ }\bibfield  {title} {\bibinfo {title} {Jupyter: {{Thinking}}
  and {{Storytelling With Code}} and {{Data}}},\ }\href
  {https://doi.org/10.1109/MCSE.2021.3059263} {\bibfield  {journal} {\bibinfo
  {journal} {Computing in Science Engineering}\ }\textbf {\bibinfo {volume}
  {23}},\ \bibinfo {pages} {7} (\bibinfo {year} {2021})}\BibitemShut {NoStop}%
\bibitem [{\citenamefont {Landau}\ and\ \citenamefont
  {Lifshitz}(1935)}]{Landau1935}%
  \BibitemOpen
  \bibfield  {author} {\bibinfo {author} {\bibfnamefont {L.~D.}\ \bibnamefont
  {Landau}}\ and\ \bibinfo {author} {\bibfnamefont {E.~M.}\ \bibnamefont
  {Lifshitz}},\ }\bibfield  {title} {\bibinfo {title} {On the theory of the
  dispersion of magnetic permeability in ferromagnetic bodies},\ }\href
  {https://doi.org/10.1016/B978-0-08-036364-6.50008-9} {\bibfield  {journal}
  {\bibinfo  {journal} {Physikalishe Zeitschrift der Sowjetunion}\ }\textbf
  {\bibinfo {volume} {8}},\ \bibinfo {pages} {153} (\bibinfo {year}
  {1935})}\BibitemShut {NoStop}%
\bibitem [{\citenamefont {Gilbert}(2004)}]{Gilbert2004}%
  \BibitemOpen
  \bibfield  {author} {\bibinfo {author} {\bibfnamefont {T.~L.}\ \bibnamefont
  {Gilbert}},\ }\bibfield  {title} {\bibinfo {title} {A phenomenological theory
  of damping in ferromagnetic materials},\ }\href
  {https://doi.org/10.1109/TMAG.2004.836740} {\bibfield  {journal} {\bibinfo
  {journal} {IEEE Transactions on Magnetics}\ }\textbf {\bibinfo {volume}
  {40}},\ \bibinfo {pages} {3443} (\bibinfo {year} {2004})}\BibitemShut
  {NoStop}%
\bibitem [{\citenamefont {Beg}\ \emph {et~al.}(2015)\citenamefont {Beg},
  \citenamefont {Carey}, \citenamefont {Wang}, \citenamefont
  {{Cort{\'e}s-Ortu{\~n}o}}, \citenamefont {Vousden}, \citenamefont {Bisotti},
  \citenamefont {Albert}, \citenamefont {Chernyshenko}, \citenamefont
  {Hovorka}, \citenamefont {Stamps},\ and\ \citenamefont {Fangohr}}]{Beg2015}%
  \BibitemOpen
  \bibfield  {author} {\bibinfo {author} {\bibfnamefont {M.}~\bibnamefont
  {Beg}}, \bibinfo {author} {\bibfnamefont {R.}~\bibnamefont {Carey}}, \bibinfo
  {author} {\bibfnamefont {W.}~\bibnamefont {Wang}}, \bibinfo {author}
  {\bibfnamefont {D.}~\bibnamefont {{Cort{\'e}s-Ortu{\~n}o}}}, \bibinfo
  {author} {\bibfnamefont {M.}~\bibnamefont {Vousden}}, \bibinfo {author}
  {\bibfnamefont {M.-A.}\ \bibnamefont {Bisotti}}, \bibinfo {author}
  {\bibfnamefont {M.}~\bibnamefont {Albert}}, \bibinfo {author} {\bibfnamefont
  {D.}~\bibnamefont {Chernyshenko}}, \bibinfo {author} {\bibfnamefont
  {O.}~\bibnamefont {Hovorka}}, \bibinfo {author} {\bibfnamefont {R.~L.}\
  \bibnamefont {Stamps}},\ and\ \bibinfo {author} {\bibfnamefont
  {H.}~\bibnamefont {Fangohr}},\ }\bibfield  {title} {\bibinfo {title} {Ground
  state search, hysteretic behaviour and reversal mechanism of skyrmionic
  textures in confined helimagnetic nanostructures},\ }\href
  {https://doi.org/10.1038/srep17137} {\bibfield  {journal} {\bibinfo
  {journal} {Scientific Reports}\ }\textbf {\bibinfo {volume} {5}},\ \bibinfo
  {pages} {17137} (\bibinfo {year} {2015})}\BibitemShut {NoStop}%
\bibitem [{\citenamefont {Beg}\ \emph {et~al.}(2022{\natexlab{a}})\citenamefont
  {Beg}, \citenamefont {Lang},\ and\ \citenamefont {Fangohr}}]{Beg2022}%
  \BibitemOpen
  \bibfield  {author} {\bibinfo {author} {\bibfnamefont {M.}~\bibnamefont
  {Beg}}, \bibinfo {author} {\bibfnamefont {M.}~\bibnamefont {Lang}},\ and\
  \bibinfo {author} {\bibfnamefont {H.}~\bibnamefont {Fangohr}},\ }\bibfield
  {title} {\bibinfo {title} {Ubermag: {{Toward More Effective Micromagnetic
  Workflows}}},\ }\href {https://doi.org/10.1109/TMAG.2021.3078896} {\bibfield
  {journal} {\bibinfo  {journal} {IEEE Transactions on Magnetics}\ }\textbf
  {\bibinfo {volume} {58}},\ \bibinfo {pages} {1} (\bibinfo {year}
  {2022}{\natexlab{a}})}\BibitemShut {NoStop}%
\bibitem [{\citenamefont {Beg}\ \emph {et~al.}(2017)\citenamefont {Beg},
  \citenamefont {Pepper},\ and\ \citenamefont {Fangohr}}]{Beg2017a}%
  \BibitemOpen
  \bibfield  {author} {\bibinfo {author} {\bibfnamefont {M.}~\bibnamefont
  {Beg}}, \bibinfo {author} {\bibfnamefont {R.~A.}\ \bibnamefont {Pepper}},\
  and\ \bibinfo {author} {\bibfnamefont {H.}~\bibnamefont {Fangohr}},\
  }\bibfield  {title} {\bibinfo {title} {User interfaces for computational
  science: {{A}} domain specific language for {{OOMMF}} embedded in
  {{Python}}},\ }\href {https://doi.org/10.1063/1.4977225} {\bibfield
  {journal} {\bibinfo  {journal} {AIP Advances}\ }\textbf {\bibinfo {volume}
  {7}},\ \bibinfo {pages} {056025} (\bibinfo {year} {2017})}\BibitemShut
  {NoStop}%
\bibitem [{\citenamefont {Beg}\ \emph {et~al.}(2022{\natexlab{b}})\citenamefont
  {Beg}, \citenamefont {Lang}, \citenamefont {Fangohr},\ and\ \citenamefont
  {Leliaert}}]{Beg2022a}%
  \BibitemOpen
  \bibfield  {author} {\bibinfo {author} {\bibfnamefont {M.}~\bibnamefont
  {Beg}}, \bibinfo {author} {\bibfnamefont {M.}~\bibnamefont {Lang}}, \bibinfo
  {author} {\bibfnamefont {H.}~\bibnamefont {Fangohr}},\ and\ \bibinfo {author}
  {\bibfnamefont {J.}~\bibnamefont {Leliaert}},\ }\href
  {https://doi.org/10.5281/zenodo.6219083} {\bibinfo {title} {Ubermag}},\
  \bibinfo {howpublished} {Zenodo} (\bibinfo {year}
  {2022}{\natexlab{b}})\BibitemShut {NoStop}%
\bibitem [{\citenamefont {Donahue}\ and\ \citenamefont
  {Porter}(1999)}]{Donahue1999}%
  \BibitemOpen
  \bibfield  {author} {\bibinfo {author} {\bibfnamefont {M.~J.}\ \bibnamefont
  {Donahue}}\ and\ \bibinfo {author} {\bibfnamefont {D.}~\bibnamefont
  {Porter}},\ }\bibfield  {title} {\bibinfo {title} {{{OOMMF User}}'s
  {{Guide}}, {{Version}} 1.0},\ }\bibfield  {journal} {\bibinfo  {journal}
  {Interagency Report NISTIR 6376, National Institute of Standards and
  Technology, Gaithersburg, MD}\ }\href {https://doi.org/10.6028/NIST.IR.6376}
  {10.6028/NIST.IR.6376} (\bibinfo {year} {1999})\BibitemShut {NoStop}%
\bibitem [{\citenamefont {{Cort{\'e}s-Ortu{\~n}o}}\ \emph
  {et~al.}(2018{\natexlab{a}})\citenamefont {{Cort{\'e}s-Ortu{\~n}o}},
  \citenamefont {Beg}, \citenamefont {Nehruji}, \citenamefont {Breth},
  \citenamefont {Pepper}, \citenamefont {Kluyver}, \citenamefont {Downing},
  \citenamefont {Hesjedal}, \citenamefont {Hatton}, \citenamefont {Lancaster},
  \citenamefont {Hertel}, \citenamefont {Hovorka},\ and\ \citenamefont
  {Fangohr}}]{Cortes-Ortuno2018}%
  \BibitemOpen
  \bibfield  {author} {\bibinfo {author} {\bibfnamefont {D.}~\bibnamefont
  {{Cort{\'e}s-Ortu{\~n}o}}}, \bibinfo {author} {\bibfnamefont
  {M.}~\bibnamefont {Beg}}, \bibinfo {author} {\bibfnamefont {V.}~\bibnamefont
  {Nehruji}}, \bibinfo {author} {\bibfnamefont {L.}~\bibnamefont {Breth}},
  \bibinfo {author} {\bibfnamefont {R.}~\bibnamefont {Pepper}}, \bibinfo
  {author} {\bibfnamefont {T.}~\bibnamefont {Kluyver}}, \bibinfo {author}
  {\bibfnamefont {G.}~\bibnamefont {Downing}}, \bibinfo {author} {\bibfnamefont
  {T.}~\bibnamefont {Hesjedal}}, \bibinfo {author} {\bibfnamefont
  {P.}~\bibnamefont {Hatton}}, \bibinfo {author} {\bibfnamefont
  {T.}~\bibnamefont {Lancaster}}, \bibinfo {author} {\bibfnamefont
  {R.}~\bibnamefont {Hertel}}, \bibinfo {author} {\bibfnamefont
  {O.}~\bibnamefont {Hovorka}},\ and\ \bibinfo {author} {\bibfnamefont
  {H.}~\bibnamefont {Fangohr}},\ }\bibfield  {title} {\bibinfo {title}
  {Proposal for a micromagnetic standard problem for materials with
  {{Dzyaloshinskii}}\textendash{{Moriya}} interaction},\ }\href
  {https://doi.org/10.1088/1367-2630/aaea1c} {\bibfield  {journal} {\bibinfo
  {journal} {New Journal of Physics}\ }\textbf {\bibinfo {volume} {20}},\
  \bibinfo {pages} {113015} (\bibinfo {year} {2018}{\natexlab{a}})}\BibitemShut
  {NoStop}%
\bibitem [{\citenamefont {{Cort{\'e}s-Ortu{\~n}o}}\ \emph
  {et~al.}(2018{\natexlab{b}})\citenamefont {{Cort{\'e}s-Ortu{\~n}o}},
  \citenamefont {Beg}, \citenamefont {Nehruji}, \citenamefont {Pepper},\ and\
  \citenamefont {Fangohr}}]{Cortes-Ortuno2018a}%
  \BibitemOpen
  \bibfield  {author} {\bibinfo {author} {\bibfnamefont {D.}~\bibnamefont
  {{Cort{\'e}s-Ortu{\~n}o}}}, \bibinfo {author} {\bibfnamefont
  {M.}~\bibnamefont {Beg}}, \bibinfo {author} {\bibfnamefont {V.}~\bibnamefont
  {Nehruji}}, \bibinfo {author} {\bibfnamefont {R.~A.}\ \bibnamefont
  {Pepper}},\ and\ \bibinfo {author} {\bibfnamefont {H.}~\bibnamefont
  {Fangohr}},\ }\href {https://doi.org/10.5281/zenodo.1196820} {\bibinfo
  {title} {{{OOMMF}} extension: {{Dzyaloshinskii-Moriya}} interaction ({{DMI}})
  for crystallographic classes {{T}} and {{O}}}},\ \bibinfo {howpublished}
  {Zenodo} (\bibinfo {year} {2018}{\natexlab{b}})\BibitemShut {NoStop}%
\bibitem [{\citenamefont {Im}\ \emph {et~al.}(2019)\citenamefont {Im},
  \citenamefont {Han}, \citenamefont {Jung}, \citenamefont {Yu}, \citenamefont
  {Lee}, \citenamefont {Yoon}, \citenamefont {Chao}, \citenamefont {Fischer},
  \citenamefont {Hong},\ and\ \citenamefont {Lee}}]{Im2019}%
  \BibitemOpen
  \bibfield  {author} {\bibinfo {author} {\bibfnamefont {M.-Y.}\ \bibnamefont
  {Im}}, \bibinfo {author} {\bibfnamefont {H.-S.}\ \bibnamefont {Han}},
  \bibinfo {author} {\bibfnamefont {M.-S.}\ \bibnamefont {Jung}}, \bibinfo
  {author} {\bibfnamefont {Y.-S.}\ \bibnamefont {Yu}}, \bibinfo {author}
  {\bibfnamefont {S.}~\bibnamefont {Lee}}, \bibinfo {author} {\bibfnamefont
  {S.}~\bibnamefont {Yoon}}, \bibinfo {author} {\bibfnamefont {W.}~\bibnamefont
  {Chao}}, \bibinfo {author} {\bibfnamefont {P.}~\bibnamefont {Fischer}},
  \bibinfo {author} {\bibfnamefont {J.-I.}\ \bibnamefont {Hong}},\ and\
  \bibinfo {author} {\bibfnamefont {K.-S.}\ \bibnamefont {Lee}},\ }\bibfield
  {title} {\bibinfo {title} {Dynamics of the {{Bloch}} point in an asymmetric
  permalloy disk},\ }\href {https://doi.org/10.1038/s41467-019-08327-6}
  {\bibfield  {journal} {\bibinfo  {journal} {Nature Communications}\ }\textbf
  {\bibinfo {volume} {10}},\ \bibinfo {pages} {593} (\bibinfo {year}
  {2019})}\BibitemShut {NoStop}%
\bibitem [{\citenamefont {Volovik}(1987)}]{Volovik1987}%
  \BibitemOpen
  \bibfield  {author} {\bibinfo {author} {\bibfnamefont {G.~E.}\ \bibnamefont
  {Volovik}},\ }\bibfield  {title} {\bibinfo {title} {Linear momentum in
  ferromagnets},\ }\href {https://doi.org/10.1088/0022-3719/20/7/003}
  {\bibfield  {journal} {\bibinfo  {journal} {Journal of Physics C: Solid State
  Physics}\ }\textbf {\bibinfo {volume} {20}},\ \bibinfo {pages} {L83}
  (\bibinfo {year} {1987})}\BibitemShut {NoStop}%
\bibitem [{\citenamefont {Liu}\ \emph {et~al.}(2018)\citenamefont {Liu},
  \citenamefont {Lake},\ and\ \citenamefont {Zang}}]{Liu2018}%
  \BibitemOpen
  \bibfield  {author} {\bibinfo {author} {\bibfnamefont {Y.}~\bibnamefont
  {Liu}}, \bibinfo {author} {\bibfnamefont {R.~K.}\ \bibnamefont {Lake}},\ and\
  \bibinfo {author} {\bibfnamefont {J.}~\bibnamefont {Zang}},\ }\bibfield
  {title} {\bibinfo {title} {Binding a hopfion in a chiral magnet nanodisk},\
  }\href {https://doi.org/10.1103/PhysRevB.98.174437} {\bibfield  {journal}
  {\bibinfo  {journal} {Physical Review B}\ }\textbf {\bibinfo {volume} {98}},\
  \bibinfo {pages} {174437} (\bibinfo {year} {2018})},\ \Eprint
  {https://arxiv.org/abs/1806.01682} {arXiv:1806.01682} \BibitemShut {NoStop}%
\end{thebibliography}%


\begin{thebibliography}{0}%
\makeatletter
\providecommand \@ifxundefined [1]{%
 \@ifx{#1\undefined}
}%
\providecommand \@ifnum [1]{%
 \ifnum #1\expandafter \@firstoftwo
 \else \expandafter \@secondoftwo
 \fi
}%
\providecommand \@ifx [1]{%
 \ifx #1\expandafter \@firstoftwo
 \else \expandafter \@secondoftwo
 \fi
}%
\providecommand \natexlab [1]{#1}%
\providecommand \enquote  [1]{``#1''}%
\providecommand \bibnamefont  [1]{#1}%
\providecommand \bibfnamefont [1]{#1}%
\providecommand \citenamefont [1]{#1}%
\providecommand \href@noop [0]{\@secondoftwo}%
\providecommand \href [0]{\begingroup \@sanitize@url \@href}%
\providecommand \@href[1]{\@@startlink{#1}\@@href}%
\providecommand \@@href[1]{\endgroup#1\@@endlink}%
\providecommand \@sanitize@url [0]{\catcode `\\12\catcode `\$12\catcode
  `\&12\catcode `\#12\catcode `\^12\catcode `\_12\catcode `\%12\relax}%
\providecommand \@@startlink[1]{}%
\providecommand \@@endlink[0]{}%
\providecommand \url  [0]{\begingroup\@sanitize@url \@url }%
\providecommand \@url [1]{\endgroup\@href {#1}{\urlprefix }}%
\providecommand \urlprefix  [0]{URL }%
\providecommand \Eprint [0]{\href }%
\providecommand \doibase [0]{https://doi.org/}%
\providecommand \selectlanguage [0]{\@gobble}%
\providecommand \bibinfo  [0]{\@secondoftwo}%
\providecommand \bibfield  [0]{\@secondoftwo}%
\providecommand \translation [1]{[#1]}%
\providecommand \BibitemOpen [0]{}%
\providecommand \bibitemStop [0]{}%
\providecommand \bibitemNoStop [0]{.\EOS\space}%
\providecommand \EOS [0]{\spacefactor3000\relax}%
\providecommand \BibitemShut  [1]{\csname bibitem#1\endcsname}%
\let\auto@bib@innerbib\@empty
\end{thebibliography}%

\section{Acknowledgements}
This work was financially supported by the EPSRC Programme grant on Skyrmionics
(EP/N032128/1). We acknowledge the use of the IRIDIS High Performance Computing
Facility, and associated support services at the University of Southampton, and
the HPC system at the Max Planck Institute for Structure and Dynamics of Matter,
in the completion of this work.

\section{Author contributions}
M.L., M.B. and H.F. conceived the study, and M.L. performed finite-difference
micromagnetic simulations and analysed the data. M.L., M.B., and H.F. developed
the Ubermag software package to drive the finite-difference solver OOMMF. M.L.,
with the assistance of H.F., M.B., and O.H. prepared the manuscript.

\section{Competing interests}
The authors declare no competing interests.

\end{document}


\title{Bloch points in nanostrips -- Supplementary material}

\author{Martin Lang}

\author{Marijan Beg}

\author{Ondrej Hovorka}

\author{Hans Fangohr}

\maketitle
\section{Initialisation and energy minimisation scheme}

In the main text we have explained that we divide the nanostrip into multiple
subregions to initialise the system and enforce the formation of a specific
number of \BP{}s during the energy minimisation. Here, we show two different
examples for configurations containing one and two \BP{}s, respectively.

\begin{figure}[ht]
  \centering
  \includegraphics[width=.8\linewidth]{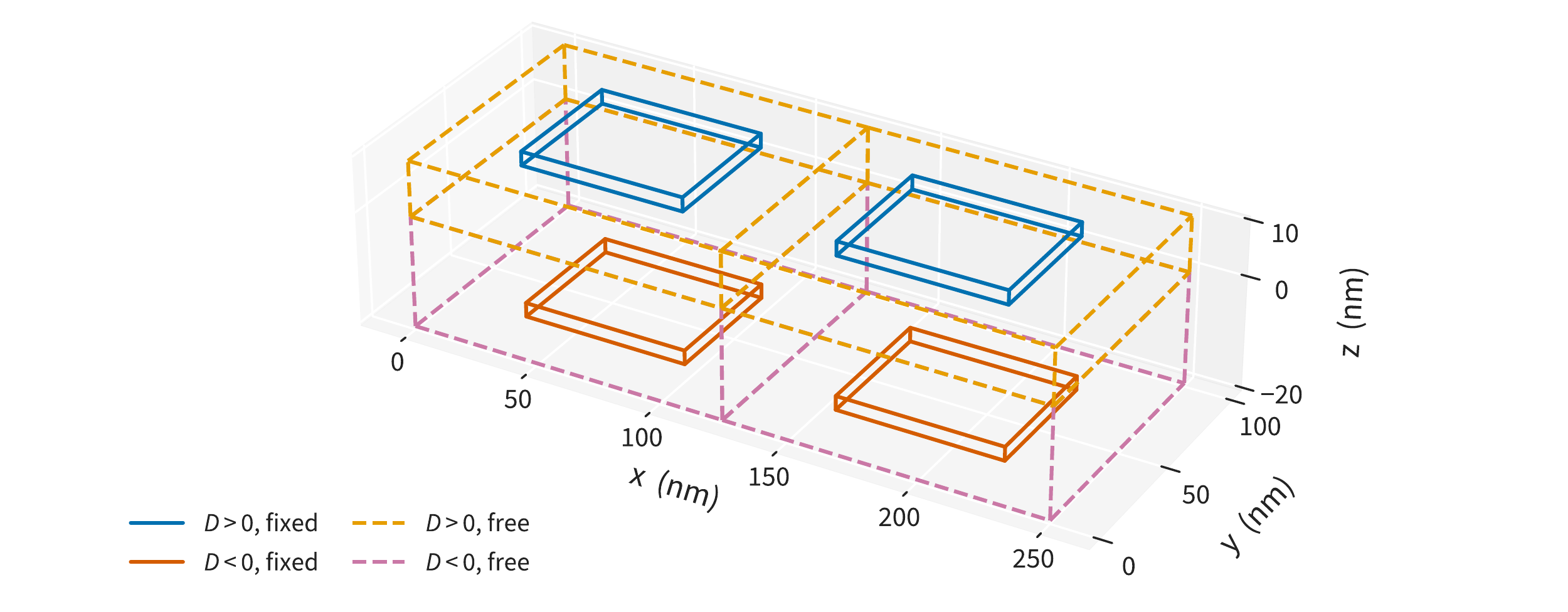}
  \caption{\label{sup:fig:subregions}Subregions used for the initialisation and
    fixed energy minimisation, and to define the two-layer geometry with
    opposite chirality (opposite sign of $D$) in the two layers for a nanostrip
    that shall contain two Bloch points. The magnetisation inside the subregions
    shown with solid lines is kept fixed during the first energy minimisation.
    Magnetisation in the top fixed subregions is initialised with reversed $z$
    component of the magnetisation (see Fig.~\ref{sup:fig:minimisation-scheme}).
    In the plot the $z$ axis is stretched for better visibility. The fixed
    subregions are located at the strip boundary.}
\end{figure}

Figure~\ref{sup:fig:subregions} shows the eight subregions that are used to
obtain a configuration containing two \BP{}s. Four subregions are located below
$z=0$ and have a negative DM energy constant $D$, and four above $z=0$ with
positive $D$. The four small subregions, shown with solid lines, are located at
the top and bottom sample boundary. Each of the small subregions is contained
within one surrounding subregions shown with dashed lines. The strip geometry in
Fig.~\ref{sup:fig:subregions} is stretched in the $z$ direction for better
visibility.

\begin{figure}[h]
\includegraphics[width=\linewidth]{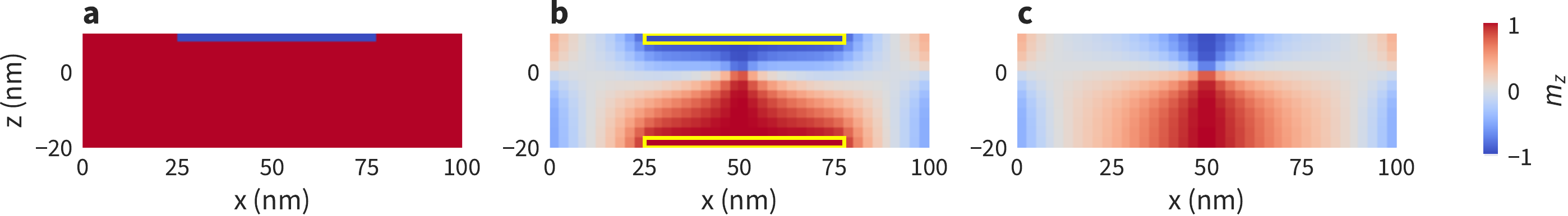}
\caption{\label{sup:fig:minimisation-scheme}Initialisation and energy
  minimsation for a single head-to-head Bloch point (cross-section at
  $y=50\,\mathrm{nm}$). (a) The magnetisation is initialised with
  $\mathbf{m}=(0, 0, 1)$ with the exeption of a top subregion where
  $\mathbf{m}=(0, 0, -1)$. (b) During the first energy minimisation the
  magnetisation is kept fixed inside the yellow-highlighted subregions (see
  Fig.~\ref{sup:fig:subregions} for a 3D plot). The formation of a \BP{} is
  enforced by the opposite $m_z$ values in the two fixed subregions. (c) During
  the second energy minimisation magnetisation in all cells can freely change.}
\end{figure}

As outlined in the main text, the system initialisation and energy minimisation
is done in three steps. The magnetisation after each step is show in
Fig.~\ref{sup:fig:minimisation-scheme} for a configuration containing a single
HH \BP{}. The system geometry is $100\,\mathrm{nm} \times 100\,\mathrm{nm}
\times (10 + 20)\,\mathrm{nm}$. This system only needs four subregions,
\emph{i.e.} one ``column'' of subregions in Fig.~\ref{sup:fig:subregions}
(\emph{e.g.} only the subregions for $x < 125\,\mathrm{nm}$).

During the initialisation step the magnetisation is initialised uniformly
throughout the sample with $\mathbf{m} = (0, 0, 1)$. The only exception is the
small subregion at the top sample boundary where the $z$ component of the
magnetisation is reversed, $\mathbf{m} = (0, 0, -1)$
(Fig.~\ref{sup:fig:minimisation-scheme}a). During the first energy minimisation
the magnetisation in the small subregions at the sample boundary, shown with
solid lines in Fig.~\ref{sup:fig:subregions} and highlighted in
Fig.~\ref{sup:fig:minimisation-scheme}b, is fixed. This enforces the formation
of the \BP{} in a controlled manner. The first energy minimisation leads to the
magnetisation configuration shown in Fig.~\ref{sup:fig:minimisation-scheme}b.
During the second energy minimisation, the magnetisation in all cells can freely
change to allow the system find the local energy minimum, leading to the
magnetisation shown in Fig.~\ref{sup:fig:minimisation-scheme}c.

\section{Classification accuracy}

In the methods section of the main text we have introduced our classification
method. In short, to classify nanostrips that contain multiple \BP{}s we compute
the convolution of the divergence of the emergent magnetic field $\mathbf{F}$
with a Heaviside step function $\Theta$:
\begin{equation}
  S(x) = \frac{1}{4\pi} \int_{V'}\mathrm{d}^3r' \, \Theta(x - x') \nabla_{\mathbf{r}'} \cdot \mathbf{F}(\mathbf{r}'). \label{eq:S-of-x}
\end{equation}
The components of $\mathbf{F}$ are defined as:
\begin{equation}
  F_i = \mathbf{m} \cdot \left( \partial_j \mathbf{m} \times \partial_k \mathbf{m} \right),
\end{equation}
where $(i, j, k)$ is an even permutation of $(x, y, z)$. The integral generally
has non-integer values due to numerical inaccuracies resulting mainly from the
finite discretisation cell size. To simplify the classification and counting we
round $S(x)$ to integer values whereby we obtain sharp steps ($\Delta S = \pm
1$) at \BP[-]{} positions.

\begin{figure}[h]
  \includegraphics[width=\linewidth]{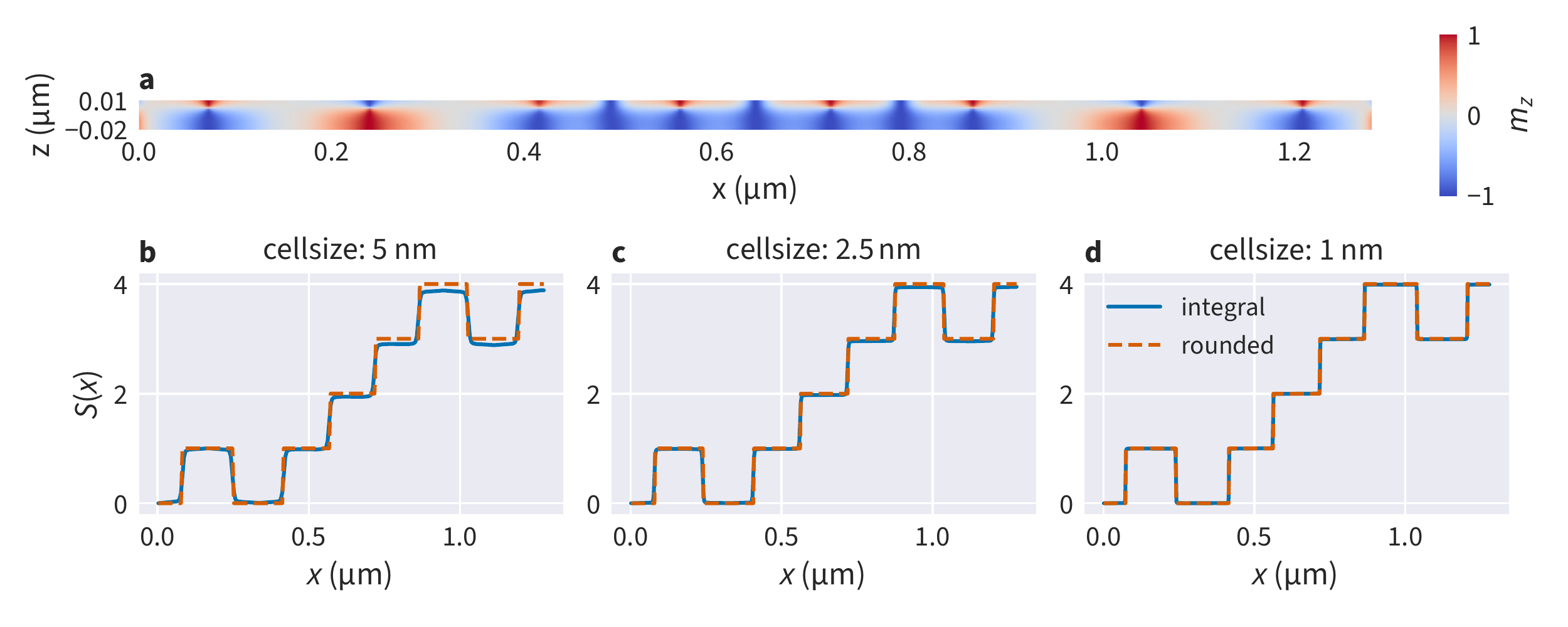}
  \caption{\label{sup:fig:classification-accuracy}Accuracy of the classification
    for a configuration containing eight \BP{}s. (a) shows a cross-section of
    the magnetisation of configuration TT-HH-TT-TT-TT-TT-HH-TT in the $xz$ plane
    located at the \BP[-]{} position ($y=50\,\mathrm{nm}$). With decreasing
    cellsize -- cell edge lengths: $5\,\mathrm{nm}$~(b), $2.5\,\mathrm{nm}$~(c),
    and $1\,\mathrm{nm}$~(d) -- the difference between the integral $S(x)$ and
    the corresponding integer values decreases.}
\end{figure}

To justify the rounding to integer values we compare the deviations of $S(x)$
from integer values for different cell sizes.
Figure~\ref{sup:fig:classification-accuracy}a shows a configuration containing
eight \BP{}s in the pattern TT-HH-TT-TT-TT-TT-HH-TT. We compute $S(x)$ for three
different cell sizes with cubic cells with edge lengths $l_{\mathrm{c}} =
5\,\mathrm{nm}$ (Fig.~\ref{sup:fig:classification-accuracy}b), $l_{\mathrm{c}} =
2.5\,\mathrm{nm}$ (Fig.~\ref{sup:fig:classification-accuracy}c), and
$l_{\mathrm{c}} = 1\,\mathrm{nm}$ (Fig.~\ref{sup:fig:classification-accuracy}d).
For each cell size we show the integral result (solid lines) and rounded values
(dashed lines). We can see that the difference (\emph{i.e.} the inaccuracy)
significantly decreases with decreasing cell size. For $l_{\mathrm{c}} =
1\,\mathrm{nm}$ integral and rounded values cannot be distinguished visually.
All simulations in the paper are done with a cell edge length $l_{\mathrm{c}} =
2.5\,\mathrm{nm}$.

